\newif\ifSTCO
\STCOfalse
\ifSTCO
  \documentclass[twocolumn, final]{svjour3}[natbib]
\else
  \documentclass[a4paper, 12pt]{article}          
\fi

\usepackage{graphicx}
\usepackage{hyperref}
\usepackage{amssymb,amsmath}
\usepackage{amsbsy}
\usepackage{bm}
\usepackage{float}
\ifSTCO \else \usepackage[margin=1in]{geometry}\fi
\usepackage{color}
\usepackage[authoryear,semicolon]{natbib}
\ifSTCO \else \usepackage{setspace} \fi
\usepackage{multirow}
\usepackage{booktabs}
\usepackage{tikz}
\usepackage{array}
\usepackage[normalem]{ulem}
\usepackage{centernot}
\usepackage{authblk}
\usepackage{comment}
\usetikzlibrary{backgrounds}
\usepackage[all]{xy}
\usepackage{authblk}
\usepackage{subcaption}
\usepackage{framed}
\newcounter{algorithm}
\newenvironment{algorithm}[1][]{\refstepcounter{algorithm}\par\medskip\noindent%
  \textbf{Algorithm~\thealgorithm: #1} \rmfamily}{\medskip}

\newcommand{\curly}[1]{\mathcal{#1}}
\newcommand{\curlyF}{{\curly{F}}}

\newcommand{\curlyS}{{\curly{S}}}
\newcommand{\curlyT}{{\curly{T}}}

\newcommand{\curlyI}{{\curly{I}}}

\newcolumntype{x}[1]{>{\centering\arraybackslash\hspace{0pt}}p{#1}}

\ifSTCO
\newcommand{\change}{\blue}
\else
\newcommand{\change}{}
\fi

\renewcommand{\epsilon}{\varepsilon}

\newcommand{\zerob} {{\bf 0}}

\newcommand{\thetab} {{\boldsymbol{\theta}}}

\newcommand{\nub} {{\boldsymbol{\nub}}}

\newcommand{\etab} {{\boldsymbol{\eta}}}
\newcommand{\Thetab} {{\boldsymbol{\Theta}}}

\newcommand{\intd} {\,\textrm{d}}

\newcommand{\Amat} {\textbf{A}}

\newcommand{\Qmat} {\textbf{Q}}

\newcommand{\Smat} {\textbf{S}}

\newcommand{\Zmat} {\textbf{Z}}
\newcommand{\Xmat} {\textbf{X}}
\newcommand{\Xvec} {\mathbf{X}}

\newcommand{\bvec} {\textbf{b}}

\newcommand{\avec} {\textbf{a}}

\newcommand{\hvec} {\textbf{h}}

\newcommand{\svec} {\textbf{s}}

\newcommand{\muvec} {\boldsymbol{\mu}}

\newcommand{\betab} {\boldsymbol {\beta}}

\renewcommand{\zerob}{\mathbf{0}}

\newcommand{\Yvec}{\mathbf{Y}}

\newcommand{\Zvec}{\mathbf{Z}}
\newcommand{\epsilonb}{\boldsymbol{\varepsilon}}

\newcommand{\E}{\mathrm{E}}

\newcommand{\var}{\mathrm{var}}

\newcommand{\Gau}{\mathrm{Gau}}

\renewcommand{\a}{\mathbf{a}}

\newcommand{\btheta}{\boldsymbol{\theta}}

\newcommand{\appropto}{\mathrel{\vcenter{
  \offinterlineskip\halign{\hfil$##$\cr
    \propto\cr\noalign{\kern2pt}\sim\cr\noalign{\kern-2pt}}}}}

\DeclareMathOperator*{\gap}{gap}

\DeclareMathOperator*{\supp}{supp}
\DeclareMathOperator{\pr}{\mathrm{pr}}

\DeclareMathOperator{\Gammaop}{\mathrm{\Gamma}}
\DeclareMathOperator{\Deltaop}{\mathrm{\Delta}}

\newcommand{\eelse}{\mathrm{eelse}}
\newcommand{\excl}{\mathrm{excl.}}
\newcommand{\rest}{\mathrm{rest}}

\newcommand{\given}{\mathbin{\vert}\nolinebreak}
\newcommand{\Perp}{\ensuremath{\mathrel{\perp\!\!\perp}}\nolinebreak}

\newcommand{\IID}{\buildrel \rm IID \over \sim}
\newcommand{\eff}{\textrm{eff}}

\let\originalleft\left
\let\originalright\right
\renewcommand{\left}{\mathopen{}\mathclose\bgroup\originalleft}
\renewcommand{\right}{\aftergroup\egroup\originalright}
\renewcommand{\tilde}{\widetilde}

\newcommand*\mathinhead[2]{\texorpdfstring{$#1$}{$#2$}}

\newfloat{algorithm-float}{thp}

\begin{document}

\sloppy

\title{\ifSTCO \else \vspace{-0.5in} \fi Multi-Scale Process Modelling and Distributed Computation for Spatial Data} 

\ifSTCO

\author{Andrew Zammit-Mangion \and Jonathan Rougier}
\institute{A. Zammit-Mangion \at
           School of Mathematics and Applied Statistics, \\
           University of Wollongong, Australia \\
              Tel.: +61-2-4221-5112\\
              \email{azm@uow.edu.au}           
           \and
           J. Rougier \at
           School of Mathematics, \\
           University of Bristol, UK \\
           \email{j.c.rougier@bristol.ac.uk}}

\else

\author{Andrew Zammit-Mangion\thanks{Corresponding author: Tel.: +61-2-4221-5112; E-mail: azm@uow.edu.au}}
\affil{\vspace{-0.2in} School of Mathematics and Applied Statistics, University of Wollongong, Wollongong, Australia \\ ORCID: orcid.org/0000-0002-4164-6866}
\author{Jonathan Rougier}
\affil{\vspace{-0.2in} School of Mathematics, University of Bristol, Bristol, UK \\ ORCID: orcid.org/0000-0003-3072-7043 \\ \vspace{-0.2in}}

\fi

\ifSTCO \date{Received: date / Accepted: date} \else \date{} \fi

\ifSTCO \else \providecommand{\keywords}[1]{\vspace{0.1in} \noindent \textbf{{Keywords: }} #1} \fi

\maketitle

\begin{abstract}
  Recent years have seen a huge development in spatial modelling and prediction methodology, driven by the increased availability of remote-sensing data and the reduced cost of distributed-processing technology. It is well known that modelling and prediction using infinite-dimensional process models is not possible with large data sets, and that both approximate models and, often, approximate-inference methods, are needed. The problem of fitting simple global spatial models to large data sets has been solved through the likes of multi-resolution approximations and nearest-neighbour techniques. Here we tackle the next challenge, that of fitting complex, nonstationary, multi-scale models to large data sets. We propose doing this through the use of superpositions of spatial processes with increasing spatial scale and increasing degrees of nonstationarity. Computation is facilitated through the use of Gaussian Markov random fields and parallel Markov chain Monte Carlo based on graph colouring. The resulting model allows for both distributed computing and distributed data. Importantly, it provides opportunities for genuine model and data scaleability and yet is still able to borrow strength across large spatial scales. We illustrate a two-scale version on a data set of sea-surface temperature containing on the order of one million observations, and compare our approach to state-of-the-art spatial modelling and prediction methods.

\ifSTCO
\else
\fi

\end{abstract}

\section{Introduction}\label{sec:Intro}

Large spatial/spatio-temporal data sets are now centre-stage in several of the environmental sciences such as meteorology and glaciology. \change{Two popular} tools available to the spatial statistician to deal with such data are the hierarchical model and the closely-related notion of conditional independence \citep[][Section 2.1.5]{Cressie_2011}. In a two layer, linear, Gaussian, data-process model, the widely adopted assumption that data are conditionally independent, given \change{a low-dimensional underlying process}, is sufficient for developing inferential algorithms that scale linearly with the dimension of the data \change{(and quadratically with the dimension of the process)}. Several methods capitalise on this approach for the spatial or spatio-temporal analysis of big data; these include fixed rank kriging \citep{Cressie_2008}, predictive processes \citep{Banerjee_2008}, and a suite of approaches based on Gaussian Markov random field (GMRF) approximations to geostatistical models \cite[e.g.,][]{Rue_2002, Lindgren_2011, Nychka_2015}. For spatio-temporal variants see, for example, \citet{Sahu_2005,Dewar_2009,Cressie_2011} and \citet{Wikle_2019}.

\change{In a two-layer data-process model, dimensionality reduction of the process needs to be done with care,} as the space of the latent functions that can be reproduced is no larger than the span of the set of basis functions used for dimensionality reduction. If this space does not contain the true signal, then any residual observed signal variability will be, at best, attributed to other components in the hierarchical model, typically measurement error or fine-scale process variability. At worst, there is no such component and inferences are over-confident \citep[e.g.,][]{Zammit_2019}. \change{As such, low-dimensional models should only be used with big data if the analyst is confident that the true (hidden) process can be adequately reconstructed using the chosen basis-function representation.}

As a result, model approximations have been developed that ensure that several basis functions can be used. Two popular ones include one based on multi-scale GMRFs \citep{Nychka_2015}, and one based on a  multi-resolution approximation to the process covariance function \citep{Katzfuss_2016}. Unfortunately, models based on multi-scale GMRFs are currently hindered by the computational bottleneck of the required sparse Cholesky factorisations, while neither approach is well-suited to approximate highly nonstationary multi-scale processes. Rather, their motivation, and the associated inference methods that have been designed for them, are built on the premise that the underlying covariance function of the process is relatively simple; the approximations are made to be able to fit the model and predict with it when using large data sets. These methods work very well when this is indeed the case. The next challenge is therefore to use models and appropriate inference methods for when we have big data \emph{and} for when the underlying process is multi-scale and highly nonstationary, as is the case in many applications.

The model we propose in Section~\ref{sec:model} models the process $Y(\cdot)$ as a sum of several processes at various scales, where the degree of nonstationarity increases with the scale of the process. We approximate these processes using GMRFs so that the conditional dependence structure of the latent variables can be exploited for local processing and the use of a parallel Gibbs sampler, as discussed in Section~\ref{sec:inference}. In Section~\ref{sec:guidelines} we give some practical guidelines on how to construct the model, and in Section~\ref{sec:application} we demonstrate the two-scale variant on a data set containing on the order of one million observations using around one million basis functions, with the covariances constructed though several hundreds of parameters to allow for local nonstationarity. Section~\ref{sec:conclusion} concludes.

\section{Multi-scale nonstationary spatial processes}\label{sec:model}

In this section we detail the model through its hierarchical structure, adopting the terminology of \citet{Berliner_1996}. The top layer in the hierarchy is the data model (Section~\ref{sec:obs}), the middle layer is the process model (Section~\ref{sec:proc}), and the bottom layer is the parameter model (Section~\ref{sec:par}). We \change{often reference} subsets of vectors and matrices through superscripts. Specifically, for a vector $\Xvec$, $\Xvec^{I_1}$ denotes the sub-vector of $\Xvec$ constructed from the elements with indices in the set $I_1$. \change{Sometimes we use subscripts to reference a specific sub-vector wherever using superscript notation would be tedious; for example $\Xvec_k$ is the $k$th component, or $k$th sub-vector, of $\Xvec$, where the component's definition is context-specific. These ways of referencing can be used together; for example $\Xvec_k^{\{i\}}$ refers to the $i$th element of the $k$th component of $\Xvec$.  When conditioning we sometimes use the symbol ``$\rest$'' to denote the remainder of a sub-vector, and its precise definition is to be taken from the context. Consider, for example, the $n$-dimensional vector $\Xvec$. If we are considering the distribution of $\Xvec^{\curlyS} \mid \Xvec^\rest$, where $\curlyS \subset \{1,\dots,n\}$, then $\Xvec^\rest \equiv \Xvec^{\{1,\dots,n\}\backslash\curlyS}$.} 

\subsection{Data model}\label{sec:obs}

Assume one has measurements $\Zvec \equiv (Z_1,\dots,Z_m)^\top$ of some underlying zero-mean spatial process $Y(\cdot)$ on $D \subset \mathbb{R}^2$ or $D \subseteq \mathbb{S}^2$\change{, where $\mathbb{S}^2$ is the surface of the sphere}.  Our data model is given by $Z_l = Y(\svec_l) + \varepsilon_l$, for $l = 1,\dots,m$, where $\varepsilon_l$ is Gaussian measurement-error that \change{is independent of $Y(\cdot)$ and that} has variance $\sigma^2_\epsilon(\svec_l)$, and $\{\svec_1,\dots,\svec_m \}\subset D$ is a set of point-referenced measurement locations. In this article we assume that $\{\varepsilon_l\}$ are uncorrelated conditional on $\sigma^2_\epsilon(\cdot)$, and that \change{$m$} is large; in our case study in Section~\ref{sec:application}, $m$ is on the order of one million.

\subsection{Process model}\label{sec:proc}

We define $Y(\cdot)$ to be the sum of $(K+1)$ independent Gaussian processes (GPs), $Y(\cdot) = Y_0(\cdot) + \sum_{k=1}^K Y_k(\cdot),$ where $Y_0(\cdot)$ is stationary with \change{large spatial scale (i.e., a large range)}, and $Y_k(\cdot), k \ge 1,$ is nonstationary with spatial scale decreasing with increasing $k$. In our approach $K \ge 1$, and in the case study of Section~\ref{sec:application}, $K = 1$.

Without loss of generality we assume that $Y_k(\cdot), k = 0,\dots,K,$ have zero expectation. We model the GPs through the stochastic partial differential equations
\begin{equation}
  (\kappa_0^2 - \Deltaop)^{\alpha_0 / 2}(\tau_0Y_0(\svec)) = W_0(\svec),\quad \svec \in D, \label{eq:SPDE0}
\end{equation}
\change{and, for $k = 1,\dots, K$},
\begin{equation}
  (\kappa_k^2(\svec) - \Deltaop)^{\alpha_k / 2}(\tau_k(\svec)Y_k(\svec)) = W_k(\svec), \quad \svec \in D,\label{eq:SPDE1}
\end{equation}
where $W_k(\cdot), k = 0,\dots,K,$ are Gaussian spatial white noise processes, $\kappa_k$ are spatial scale parameters, $\tau_k$ control the process variance, and $\Deltaop$ is the Laplacian. When $D \subset \mathbb{R}^2$, the solution to \eqref{eq:SPDE0} is a stationary GP with Mat{\'e}rn covariance function with smoothness parameter $\nu_0 = \alpha_0 - 1$, while that to \eqref{eq:SPDE1} is a nonstationary process which, under some regularity assumptions on $\kappa_k(\cdot)$ and $\tau_k(\cdot)$,  exhibits \change{local Mat{\'e}rn behaviour with smoothness parameter $\nu_k = \alpha_k - 1$ \citep{Lindgren_2011}}. \change{Throughout, we assume that $\nu_k, k = 0,\dots,K,$ are known.}

As in \citet{Lindgren_2011} and \citet{Lindgren_2015}, we project $Y_k(\cdot), k = 0, 1,\dots,K,$ onto a finite-dimensional basis through $Y_k(\cdot) = \avec_{k}(\cdot)^\top\etab_k$, where $\avec_k(\cdot)$ is a vector of $r_{\eta_k}$ tent basis functions on a triangulation of $D$, $T_{\eta_k}$ say \change{\citep[][Section 12.6.2]{Scherer_2017}}, and $\etab_k \sim \Gau(\zerob, \Qmat_k^{-1})$, where $\Qmat_k$ is a sparse precision matrix. \change{(Throughout, we use $T_{\beta}$ to denote the triangulation associated with the random quantity $\betab$).} \change{The triangulations $T_{\eta_k}, k = 0,\dots,K,$ become finer with increasing $k$ since $Y_k(\cdot)$ has spatial scale that decreases with $k$.} \change{Since each basis function corresponds to a vertex in the triangulation, t}he coefficients $\etab_k, k = 0,\dots, K,$\change{~which are independent across $k$,} can be associated with the vertices of the basis functions they weight; this property will facilitate graph partitioning and colouring in Section \ref{sec:inference}.

After projecting onto our finite-dimensional subspace, we can write the measurements as 
$$
\Zvec = \sum_{k = 0}^K\Amat_{k}\etab_k + \epsilonb,
$$
where $\Amat_k \equiv (\avec_k(\svec_l): l = 1,\dots,m)^\top$, $\epsilonb \sim \Gau(\zerob, \Qmat_\epsilon^{-1}),$ and  $\Qmat_\epsilon$ is diagonal with measurement-error precisions as non-zero elements.  
Collect all the unknown model parameters \change{used to construct $\Qmat_\epsilon$ and $\Qmat \equiv \textrm{bdiag}(\Qmat_0,\dots,\Qmat_K)$} into the vector $\thetab$. Let $\Amat\equiv (\Amat_0, \dots,\Amat_K)$ and $\etab \equiv (\etab_0^\top, \dots,\etab_{K}^\top)^\top$. For this model, which is structurally identical to the LatticeKrig model \citep{Nychka_2015}, 
\change{$\etab \mid \Zvec , \thetab$ also has a sparse precision matrix since $\Zvec$ is point referenced}. However, both $\Zvec$ and $\etab$ can contain millions of elements so that storing, let alone computing with, 
\change{$\Amat$ and this precision matrix,} can become impractical. On the other hand GMRFs have the desirable property that they admit parallel-data and parallel-model conditional structures that can be used to facilitate computation. More details on how we exploit this property are given in Section~\ref{sec:inference}.

\subsection{Parameter model}\label{sec:par}

As in \citet{Lindgren_2015} we do not consider the natural parameters \change{$\tau_0, \kappa_0$, and $\tau_k(\cdot), \kappa_k(\cdot), k = 1,\dots,K,$ directly. Rather, we construct a parameter model for the process standard deviations $\sigma_0$ and $\{\sigma_k(\cdot)\}$, and the nominal ranges $\rho_0$ and $\{\rho_k(\cdot)\}$. In our setting, for $k = 1,\dots,K$ (and similarly for $k = 0$),  these are related to the natural parameters through} 
\change{\begin{align*}
  \log{\tau_k(\cdot)} &= \frac{1}{2}\log\left(\frac{\Gammaop(\nu_k)}{4\pi\Gammaop(\alpha_k)}\right) - \log\sigma_k(\cdot) - \nu_k\log\kappa_k(\cdot), \\
  \log{\kappa_k(\cdot)} &= \frac{1}{2}\log{(8\nu_k)} - \log\rho_k(\cdot),
\end{align*}}
\noindent \hspace{-0.1in} where $\Gammaop(\cdot)$ is the gamma function. 

We decompose the spatially-varying parameter \change{$\log \sigma_\epsilon(\cdot)$} as a weighted sum of $r_\epsilon$ \change{tent} basis functions $\bvec_\epsilon(\cdot)$  constructed on a triangulation $T_\epsilon$ on $D$. We also decompose \change{$\log \sigma_k(\cdot)$} and \change{$\log\rho_k(\cdot), k =1,\dots,K,$}  as weighted sums of \change{$r_{\theta_k} \equiv r_{\sigma_k} = r_{\rho_k}$} \change{tent} basis functions, $\bvec_{k}(\cdot)$, constructed on a triangulation \change{$T_{\theta_k} \equiv T_{\sigma_k} = T_{\rho_k}$} on $D$. \change{Note that here, and below, we use the subscript $\theta_k$ to identify quantities such as basis functions and triangulations that are common to both $\sigma_k(\cdot)$ and $\rho_k(\cdot)$.} \change{Since the parameter fields at scale $k$ need to vary slowly relative to the process at that scale \citep{Lindgren_2011}, t}he triangulation $T_{\theta_k}$ can be much coarser than $T_{\eta_k}$. These decompositions yield the following parameter models,
\change{\begin{align*}
  \log\sigma_\epsilon(\cdot) &= \bvec_{\epsilon}(\cdot)^\top\thetab_\epsilon, \\
  \log\sigma_k(\cdot) &= \bvec_k(\cdot)^\top\thetab_{\sigma_k}, \\
  \log\rho_k(\cdot) &= \bvec_k(\cdot)^\top\thetab_{\rho_k},
\end{align*}}
for $k = 1,\dots,K$, remembering that, for $k = 0$, $\sigma_0$ and $\rho_0$ are spatially-invariant.

We assume that the elements of the parameter \change{vectors $\thetab_X, X \in \{\epsilon, \sigma_1, \rho_1,\dots, \sigma_K, \rho_K\}$, are independent and distributed according to Gaussian distributions. Specifically, we let
  \begin{equation}
  \thetab_{X}^{\{i\}} \mid \omega_{X}, \lambda_{X} \IID \Gau(\omega_X, \lambda_X),\quad i = 1,\dots,r_{X},\label{eq:thetamodel1}
  \end{equation} 
  where $\omega_X = \E(\thetab_{X}^{\{i\}})$, $\lambda_X = \var(\thetab_{X}^{\{i\}})$}\change{, and $r_X$ is the dimension of $\thetab_X$}. \change{Although we assume independence, a GMRF model for the parameters might also be used \citep[see][]{Monterrubio_2019}, although this would introduce additional complexity to our updating algorithm of Section~\ref{sec:inference}.} The hyperparameters  $\omega_{\rho_k}$ and $\lambda_{\rho_k}$ need to reflect a prior judgement that the process range decreases with increasing $k$
;  we provide some guidelines in Section~\ref{sec:guidelines} and an example of how this can be done in practice with a two-scale process in Section~\ref{sec:application}. Appropriate hyperparameters for constructing the prior distributions over $\thetab_\epsilon$ and $\thetab_{\sigma_k}$ can usually be deduced from \change{domain knowledge and} exploratory data analysis \change{(see Section~\ref{sec:model_setup})}.

We complete our parameter model by equipping the parameters appearing in the coarse component, $\rho_0$ and $\sigma_0$, with the prior lognormal distributions,
\begin{align}
  \log\sigma_0 \sim \Gau(\omega_{\sigma_0}, \lambda_{\sigma_0}),\label{eq:sigma0}\\
  \log\rho_0 \sim \Gau(\omega_{\rho_0}, \lambda_{\rho_0}),\label{eq:rho0}
\end{align}
where reasonable values of $\omega_{\sigma_0}$ and $\lambda_{\sigma_0}$ can generally be deduced from exploratory analysis of the data, while $\omega_{\rho_0}$ and $\lambda_{\rho_0}$ can generally be fixed using prior application knowledge (see Sections~\ref{sec:guidelines} and~\ref{sec:application}). \change{As with $k > 0$, we assume that $\sigma_0$ and $\rho_0$ are independent.}

To summarise, the full set of parameters and the full conditional dependence structure of our model is given by the following directed acyclic graph (DAG):
\begin{equation*}
  \xymatrix{
    \thetab_\epsilon  \ar[r] & \Zvec &  \\
    \etab_0 \ar[ur] & \etab_1 \ar[u] & \cdots & \etab_K \ar[ull]\\
    (\sigma_0, \rho_0) \ar[u] &  (\thetab_{\sigma_1}, \thetab_{\rho_1})
    \ar[u] & \cdots & (\thetab_{\sigma_K}, \thetab_{\rho_K}) \ar[u]
  }
\end{equation*}
where the components of $\thetab_{\sigma_k}$ and $\btheta_{\rho_k}$ are mutually independent for $k = 1,\dots, K$. The coefficients $\etab_0$ construct a stationary process with spatially-invariant standard deviation and range, $\sigma_0$ and $\rho_0$, respectively. On the other hand, $\etab_1, \dots, \etab_K,$ construct nonstationary processes with spatially-varying standard deviations and ranges.  For these nonstationary processes, the standard deviations and ranges are modelled through the coefficients  $\thetab_{\sigma_k}$ and $\thetab_{\rho_k},~k = 1,\dots,K$, where the individual elements are independent and their marginal distributions are pre-specified.  The measurement errors $\epsilonb$ are assumed to be spatially uncorrelated and are thus modelled using just the basis-function coefficients, $\thetab_\epsilon$.

\section{Inference}\label{sec:inference}

\change{Our aim is to sample from the posterior distribution $\pr(\etab_0,\dots,\etab_K, \thetab_\epsilon, \thetab_0, \dots, \thetab_K \mid \Zvec)$, where $\thetab_0 \equiv (\sigma_0, \rho_0)^\top$, and $\thetab_k \equiv (\thetab_{\sigma_k}^\top, \thetab_{\rho_k}^\top)^\top, k = 1,\dots,K$.} The value of \change{the} model is partly that it provides a flexible representation of spatial nonstationarity, and partly that \change{its posterior distribution can be sampled from} in a serial fashion using many Gibbs steps.  Additionally, the Gibbs steps for $\etab_k$ and $\thetab_k, k = 1,\dots,K,$ can be split and parallelised if they are too difficult to sample directly. We hence have a parallel MCMC scheme that can scale well with both data size and model complexity.

In Section~\ref{sec:scale0} we discuss Gibbs sampler updates for the process and parameter coefficients in the first scale; in Section~\ref{sec:scale1} the updates for the process and parameter coefficients in the other scales; in Section~\ref{sec:weights1} a re-updating strategy for the process coefficients for $k > 0$; in Section~\ref{sec:epsilon_updates} the updates for the measurement-error variance parameters; and in Section~\ref{sec:Gibbssummary} we summarise the sampler. 

\subsection{Updating the process and parameter coefficients for scale $k = 0$}\label{sec:scale0} 

The process $Y_0(\cdot)$ captures large-scale effects, and hence its triangulation $T_{\eta_0}$ can be coarse relative to $T_{\eta_k}$ for $k = 1, \dots, K$. Since $\thetab_0 \equiv (\sigma_0, \rho_0)^\top$ and $\etab_0$ can be expected to be highly correlated a posteriori \citep{Knorr_2002} we sample them jointly in the Gibbs sampler, which substantially improves the mixing of the Markov chain. Specifically, we sample from the full conditional distribution
\change{\begin{align}
    \pr(\etab_0, \thetab_0 \mid \eelse) =  &\pr(\etab_0 \mid \thetab_0, \eelse)  \nonumber \\ 
         &~~~\times \pr(\thetab_0 \mid \eelse~\excl~\etab_0),\label{eq:joint0}
\end{align}}
\noindent \hspace{-0.09in} by first sampling from $\pr(\thetab_0 \mid \eelse~\excl~\etab_0)$ and then from $\pr(\etab_0 \mid \thetab_0, \eelse)$, where we use `$\eelse$' as shorthand for `everything else.' Here, and in the rest of the article, we always sample using the most recently sampled values of the other variables.

The conditional distribution of $\thetab_0$ in \eqref{eq:joint0} is a partially collapsed distribution, where $\etab_0$ is integrated out:
\begin{align}
\pr(\thetab_0 \mid\eelse~\excl~\etab_0) &= \int \pr(\etab_0, \thetab_0 \mid \eelse) \intd \etab_0  \nonumber \\
&\propto \int \pr(\Zvec \mid \etab_0, \eelse) \nonumber \\
&~~~~~\times \pr(\etab_0 \mid \thetab_0) \intd \etab_0 \label{eq:theta0int}
\pr(\thetab_0).\end{align} 
The integrand in \eqref{eq:theta0int} is Gaussian, and therefore the integral can be evaluated analytically to give
\begin{align}
  \log \pr(\thetab_0& \mid\eelse~\excl~\etab_0) = ~ \frac{1}{2}\log|\Qmat_\epsilon| + \frac{1}{2}\log|\Qmat_0| \nonumber \\
   & - \frac{1}{2}\log|\tilde\Qmat_0| - \frac{1}{2}\tilde\Zvec^\top\Qmat_\epsilon\tilde\Zvec~+ \frac{1}{2}\tilde{\Zvec}^\top\Qmat_\epsilon\Amat_0\tilde{\muvec}_0 \nonumber \\  
  & + \log\pr(\thetab_0) + \textrm{const.}, \label{eq:theta0cond} 
\end{align}
where $\tilde\Qmat_{0} = \Amat_{0}^\top\Qmat_\epsilon\Amat_{0} + \Qmat_{0}, \tilde\muvec_{0} = \tilde\Qmat_{0}^{-1}\Amat_{0}^\top\Qmat_\epsilon\tilde\Zmat,$ and $\tilde\Zvec = \Zvec - \sum_{k > 0}\Amat_{k}\etab_k$. The triangulation $T_{\eta_0}$ must be made sufficiently coarse that factorising $\Qmat_0$, as well as algebraic operations of the form $\tilde\Qmat_0^{-1}\Xmat = \Yvec$, can be done on a single computing node. We provide more detailed guidelines on this in Section~\ref{sec:guidelines}. 
All other operations can be computed in a distributed fashion from sums and products of smaller vectors and matrices corresponding to chunks of $\Zvec, \{\Amat_k\}$, and $\{\etab_k\}$. There is therefore no theoretical limit on the data size and number of scales $K$ that will preclude sampling from this distribution in a reasonable time-frame given sufficient parallel computing resources. This is also true for all the sampling operations we outline below. The conditional distribution \eqref{eq:theta0cond} does not have a recognisable distribution in $\thetab_0$, and so this update uses a Metropolis--Hastings step.  \change{The computational complexity of this step is dominated by the sparse Cholesky factorisation of $\tilde\Qmat_0$, which is approximately $O(r_{\eta_0}^{3/2})$ flops.}

After updating $\thetab_0$ we update $\etab_0$. The full conditional distribution of $\etab_0$ is given by 
\begin{equation}\label{eq:etab0cond}
  \etab_0 \mid \eelse \sim \Gau(\tilde\muvec_{0}, \tilde\Qmat_{0}^{-1}),
\end{equation}
where $\tilde\muvec_0$ and $\tilde\Qmat_0$ have already been computed for \eqref{eq:theta0cond}. A sample of \eqref{eq:etab0cond} is therefore obtained `for free' after the update of $\thetab_0$. 

\subsection{Updating the process and parameter coefficients for scales $k = 1,\dots,K$}\label{sec:scale1}

The update of $\thetab_k \equiv (\thetab_{\sigma_k}^\top, \thetab_{\rho_k}^\top)^\top$ is more tricky than that of $\thetab_0$:  The full conditional distribution of $\thetab_k$ has the same structure as that of $\thetab_0$ given in \eqref{eq:theta0int}, but since for $k > 0$ the triangulation $T_{\eta_k}$ is fine and $r_{\eta_k}$ is large,  the elements of the integral cannot be evaluated in memory, and so a sequential strategy is necessary. 

\begin{figure}[t!]
	\begin{center}
		\includegraphics[width=\ifSTCO 0.45\else 0.6\fi\textwidth]{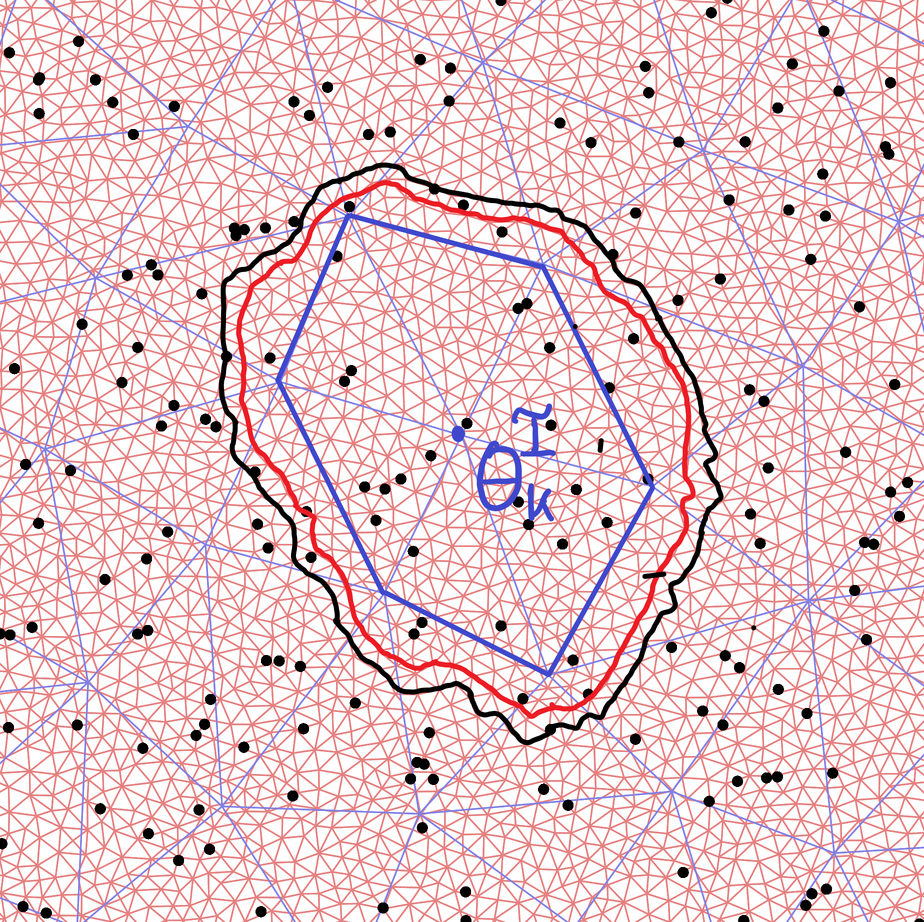}
                \caption{\change{Illustration of regions delineating the data footprint $\curlyF$ (bold black line) and the effective process footprint $\curlyT$ (bold red line) of $\thetab_k^\curlyI$ (blue dot) and the support of its associated basis function (bold blue line). The black dots depict the data points, the red mesh depicts $T_{\eta_k}$, and the blue mesh depicts $T_{\theta_k}$.}\label{fig:footprints}}
         \end{center}
\end{figure}

Let $\curlyI$ index one element of $\thetab_{\sigma_k}$ (equivalently $\thetab_{\rho_k}$) and, with a slight abuse of notation, let $\thetab_k^\curlyI$ denote the pair of elements $(\thetab_{\sigma_k}^\curlyI, \thetab_{\rho_k}^\curlyI)^\top$. Let the set $\curlyT$ be the \emph{effective process footprint} of $\thetab_k^\curlyI$, defined as the set of indices of $\etab_k$ for which the (prior) conditional distribution of $\etab_k^\rest$ \emph{excluding} $\etab_k^\curlyT$, \change{$\pr(\etab_k^\rest \given \thetab_{k}^\curlyI, \thetab_k^\rest)$}, is approximately independent of $\thetab_k^\curlyI$ \change{(see \eqref{eq:etabkmargapprox} below)}. It might be helpful to picture $\etab_k^\curlyT$ as a `halo' around $\supp(\bvec_k^{\curlyI})$, where $\supp(\cdot)$ denotes a function's support\change{; see Fig.~\ref{fig:footprints}}. Further, let $\curlyF$ be the \emph{data footprint} of $\etab_k^{\curlyT}$, that is, the set of indices of the data points that lie in the subset of the domain given by \change{$\cup_{i \in \curlyT} \supp(\avec_k^{\{i\}})$}. Under the  partitioning of the coefficients $\{\thetab_k^{\curlyI}, \thetab_k^{\rest}\}$, $\{\etab_k^\curlyT,\etab_k^\rest\}$, and the data $\{\Zvec^{\curlyF}, \Zvec^\rest\}$, the relevant part of the full DAG is
\begin{equation}\label{DAG:thetak}
\hspace{1in}      \xymatrix{
    \Zvec^{\curlyF}  & \Zvec^\rest &  \\
    \etab_k^\curlyT \ar[u] & \etab_k^\rest \ar[ul]\ar[l]\ar[u] \\
    \thetab_k^\curlyI \ar[u]\ar@{-->}[ur] &  \thetab_k^\rest \ar[ul]\ar[u]
  }
\end{equation}
\noindent where we have used a dashed line to denote a conditional dependence relationship that is weak, and one that can potentially be ignored for practical purposes.  

We jointly update $\{\etab_k^\curlyT, \thetab_k^\curlyI\}$ via the full conditional distribution
    \begin{align*}
      \pr(\etab_k^\curlyT, \thetab_k^\curlyI \mid \eelse) = &\pr(\etab_k^\curlyT \mid \thetab_k^\curlyI, \eelse) \\
      &\quad \times \pr(\thetab_k^\curlyI \mid \eelse \textrm{~excl.~} \etab_k^\curlyT),
    \end{align*}
     by first sampling from $\pr(\thetab_k^\curlyI \mid \eelse \textrm{~excl.~} \etab_k^\curlyT)$ and subsequently from $\pr(\etab_k^\curlyT \mid \thetab_k^\curlyI, \eelse)$. As with $\thetab_0$ we have that $\pr(\thetab_k^\curlyI \mid \eelse \textrm{~excl.~} \etab_k^\curlyT)$ is a partially collapsed distribution given by
         \begin{align*}
\pr(\thetab_k^\curlyI \mid &\eelse \textrm{~excl.~} \etab_k^\curlyT) = \int \pr(\thetab_k^\curlyI, \etab_k^\curlyT \mid \eelse) \intd \etab_k^\curlyT \\
      & \propto \int \pr(\Zvec^{\curlyF} \mid \etab_k^\curlyT, \eelse) \pr(\etab_k^\curlyT \mid \etab_k^\rest, \thetab_k^\curlyI, \thetab_k^\rest) \\
&~~~~~\times \pr (\etab_k^\rest \mid \thetab_k^\curlyI, \thetab_k^\rest)  \intd \etab_k^\curlyT\pr(\thetab_k^\curlyI).
    \end{align*}
Since $\etab_k$ is a Gaussian Markov random field it is computationally cheap to evaluate the conditional distribution $\pr(\etab_k^\curlyT \mid \etab_k^\rest, \thetab_k^\curlyI, \thetab_k^\rest)$. \change{However, it is computationally expensive to evaluate the `marginal' distribution $\pr(\etab_k^\rest \mid \thetab_k^\curlyI, \thetab_k^\rest)$, as is required for this update, since it has a (large) precision matrix that is not sparse in general.} We therefore resort to the approximation implied by the dashed line of the DAG in \eqref{DAG:thetak},
\begin{equation}\label{eq:etabkmargapprox}
\begin{split}
\pr(\etab_k^\rest \given \thetab_{k}^\curlyI, \thetab_k^\rest) 
\approx
\pr(\etab_k^\rest \given \tilde \thetab_k^\curlyI, \thetab_k^\rest),
\end{split}
\end{equation}
where we have replaced $\thetab_k^\curlyI$ by some value $ \tilde\thetab_{k}^\curlyI$, which may depend on the value of $\thetab_k^\rest$.  This approximation is motivated by the fact that, notionally, $\thetab_{k}^\curlyI$ controls the variance parameters $\sigma_k(\cdot)$ and $\rho_k(\cdot)$ \emph{locally}, and can therefore be expected to have diminishing effect on the probability of $\etab_k^\rest$ by the marginalisation property of the Gaussian distribution. An attractive feature of this approximation is that it can be improved as much as needed, at the expense of increased computational cost, by increasing the size of the effective process footprint $\curlyT$. The chosen effective process footprint should be one for which the approximation \eqref{eq:etabkmargapprox} is acceptable in practice. In our application, setting $\curlyT$ to be the indices of the basis functions $\avec_k(\cdot)$ that have their and their neighbours' maxima inside $\supp(\bvec_k^{\curlyI})$ sufficed, but more conservative choices could be considered if needed; see Sect.~\ref{sec:guidelines_footprint}. 

Our approximation yields
\begin{align} 
  \pr(&\thetab_k^\curlyI \mid \eelse \textrm{~excl.~} \etab_k^\curlyT) \nonumber\\
  &\appropto \int \pr(\Zvec^{\curlyF} \mid \etab_k^\curlyT, \eelse)\pr(\etab_k^\curlyT \mid \etab_k^\rest, \thetab_k^\curlyI, \thetab_k^\rest) \intd \etab_k^\curlyT \nonumber\\
 & ~~~~\times \pr(\thetab_k^\curlyI),\label{eq:thetaapprox}
\end{align}
the log of which is identical in structure to \eqref{eq:theta0cond} with an added term due to the presence of $\etab_k^\rest$ in the conditional distribution of $\etab_k^\curlyT$ \citep[][Sect.~2.2.5]{Rue_2005}. This extra term does not alter the computational or memory complexity of the operations. \change{In particular, the computational complexity is approximately $O(|\curlyT^{3/2}|)$ flops, and thus sampling will be feasible} as long as $|\curlyT|$ \change{has} the same order of magnitude as $r_{\eta_0}$, which we have set to be sufficiently small to make it possible for these computations to be done on a single computing node. 

As with $\thetab_0$, this distribution has no recognisable form and requires a Metropolis--Hastings step. Further, the computations for this conditional distribution can once again benefit from a distributed data framework. However, since there are $r_{\theta_k}$ of them (one for each parameter basis function), for the inference framework to be scaleable, the total computational time required by these updates must not depend on $r_{\theta_k}$. This is indeed the case under our approximation as we now show. 

Take two sets of parameters $\thetab_{k}^{\curlyI_1}$ and $\thetab_{k}^{\curlyI_2}$ with associated effective process footprints $\curlyT_1$ and $\curlyT_2$ for which $\etab_k^{\curlyT_1} \Perp \etab_k^{\curlyT_2} \given \etab_k^\rest, \thetab_k,$ and for which the associated data footprints $\curlyF_1 \cap \curlyF_2 = \emptyset$ (this condition is trivially satisfied in our context where $\Zvec$ are point referenced and the $\{\avec_k(\cdot)\}$ are tent basis functions).  Partition $\thetab_k$ as $\{\thetab_{k}^{\curlyI_1}, \thetab_{k}^{\curlyI_2}, \thetab_k^\rest\}$, $\etab_k$ as $\{\etab_k^{\curlyT_1}, \etab_k^{\curlyT_2}, \etab_k^\rest\}$, and $\Zvec$ as $\{\Zvec^{\curlyF_1}, \Zvec^{\curlyF_2}, \Zvec^\rest\}$. Then, the relevant part of the DAG where we now omit the weak conditional dependencies for clarity, is
  \begin{equation}\label{DAG:thetak2}
\hspace{0.3in}  \xymatrix{
    & \Zvec^{\curlyF_1}  & \Zvec^\rest & \Zvec^{\curlyF_2}   \\
    & \etab_k^{\curlyT_1} \ar[u] &  \etab_k^\rest \ar[ul]\ar[l]\ar[u]\ar[ur] \ar[r] & \etab_k^{\curlyT_2} \ar[u]  \\
    & \thetab_k^{\curlyI_1}\ar[u]    & \thetab_k^\rest \ar[ur]\ar[ul]\ar[u] &  \thetab_k^{\curlyI_2} \ar[u]\\
  }
\end{equation}
  From the moralised version \change{\citep[][Chapter 2]{Lauritzen_1996}} of \eqref{DAG:thetak2}, we immediately see that the coefficients $\{\etab_k^\rest, \thetab_k^\rest\}$ separate the two groups $\{\etab_k^{\curlyT_1}, \thetab_k^{\curlyI_1}\}$ and $\{\etab_k^{\curlyT_2}, \thetab_k^{\curlyI_2}\}$ so that 
\begin{align*}
  \pr(&\{\etab_k^{\curlyT_1}, \thetab_{k}^{\curlyI_1}\}, \{\etab_k^{\curlyT_2}, \thetab_{k}^{\curlyI_2}\} \mid \eelse) \\
  &~\appropto~ \pr(\etab_k^{\curlyT_1}, \thetab_{k}^{\curlyI_1} \mid \eelse ) \pr(\etab_k^{\curlyT_2}, \thetab_{k}^{\curlyI_2} \mid \eelse) \\
  &~=~\pr(\etab_k^{\curlyT_1} \mid \eelse)  \pr(\thetab_k^{\curlyI_1} \mid \eelse \textrm{~excl.~} \etab_k^{\curlyT_1}) \nonumber \\
  & ~~~~\times \pr(\etab_k^{\curlyT_2} \mid \eelse)\pr(\thetab_k^{\curlyI_2} \mid \eelse \textrm{~excl.~} \etab_k^{\curlyT_2}). \end{align*}
Therefore, the update operations for $\thetab_{k}^{\curlyI_1}$ and $\thetab_{k}^{\curlyI_2}$ can be dispatched to different cores in a multicore computing architecture, and the updates can be done in parallel. We can identify the sets of parameters that can be updated in parallel by \emph{colouring} the full conditional dependency graph of $\thetab_k$ under our chosen set of effective footprints, such that no conditionally dependent elements of $\thetab_k$ have the same colour. In a minimal setting, where we choose the effective process footprints as the subset of $\etab_k$ for which the maxima of the associated basis functions and those of its Markov blanket lie in the support of $\bvec_k^\curlyI$,  the conditional dependency graph is the edge-preserving bijection of the spatial graph $T_{\theta_k}$. With this choice of footprints we can then make use of the four-colour theorem \citep{Gonthier_2008} so that a sample of $\thetab_k$ can be done in exactly four steps, irrespective of the scale or data size. More conservative (i.e., larger) footprints will result in the need for more colours, and hence a lower degree of parallelisation, but still the number of colours will be independent of scale and data size, rendering this model and inferential technique scaleable. In our application we made use of the backtracking algorithm for colouring the graph \citep{Bender_1985}.

\change{Consider now the joint update of the quantities} $\{\etab_k^\curlyT, \thetab_k^\curlyI\}$.  As in the joint update of $\{\etab_0, \thetab_0\}$, a sample from the full conditional distribution of $\etab_k^\curlyT$ is obtained `for free' since it is Gaussian with a mean vector and a precision matrix that are also used when sampling $\thetab_k^\curlyI$. One might be tempted to just skip this step, obtain a full sample of $\thetab_k$, and only then obtain a full sample of $\etab_k$ using, for example, the re-updating strategy in Sect.~\ref{sec:weights1}. However, doing so will result in a collapsed Gibbs sampler that does not target the correct stationary distribution \citep{vanDyk_2008}\change{; see Appendix \ref{app:target} for more details on why this is the case}. It is therefore important that $\etab_k^\curlyT$ is sampled concurrently with $\thetab_k^\curlyI$ although, as we discuss next, a re-updating of $\etab_k$ is necessary in practice to improve convergence of the \change{Markov chain}.

\subsection{Re-updating the process coefficients for scales $k = 1,\dots,K$}\label{sec:weights1}

When a block of $\etab_k$ gets updated conditional on everything else it is pinned at its boundary, because $\etab_k$ is locally smooth a priori and, in regions where data density is low, a posteriori as well. This results in slow mixing of the \change{Markov chain}\change{, which was problematic in earlier versions of our implementation (see Appendix \ref{app:sim_blocking} for an illustration on a simple model).} We \change{now} address this issue by re-sampling each $\etab_k$ in a separate Gibbs step.

For each $\etab_k$ we tile the domain $D$ into tiles so that each element of $\etab_k$ is then associated with exactly one tile, specifically the tile in which the basis function it weights is a maximum. Each tile is no larger than what can be processed on a single node, and so will be associated with about $r_{\eta_0}$ basis functions, as per the update of $\etab_0$. The tiles must also be large enough such that the elements of $\etab_k$ associated with any two non-contiguous tiles (i.e., tiles that do not share a common boundary) are conditionally independent given the elements of $\etab_k$  associated with the other tiles. 
These tiles and their neighbours can be used to establish a supergraph made up of \emph{blocks} of $\etab_k$, where each block (corresponding to one tile) is conditionally independent of the rest given its neighbours. This supergraph is then coloured such that no two neighbouring blocks have the same colour; for point-referenced data this will require at most four colours. The colouring also corresponds to one on the tiles where no two tiles which share a common border have the same colour.

As with the joint update of $\{\etab_k, \thetab_k\}$, this colouring allows us to develop a parallel sampler for $\etab_k$. Specifically, consider two blocks of $\etab_k$ that are associated with two tiles, with indices given by $\curlyT_1$ and $\curlyT_2$, respectively. When these two blocks are of the same colour, $\etab_k^{\curlyT_1} \Perp \etab_k^{\curlyT_2} \given \etab_k^\rest$ according to the structural zeros of $\Qmat_k$.  Let $\curlyF_1$ be the data footprint of $\etab_k^{\curlyT_1}$. 
Similarly, let $\curlyF_2$ be the data footprint of $\etab_k^{\curlyT_2}$. For point-referenced measurements it is straightforward to see that $\curlyF_1 \cap \curlyF_2 = \emptyset$ so that we can partition $\Zvec$ as $\{\Zvec^{\curlyF_1}, \Zvec^{\curlyF_2}, \Zvec^\rest\}$.  The relevant part of the DAG for $\etab_k$ and $\Zvec$ is therefore simply the \change{first two lines of} \eqref{DAG:thetak2}, from which we obtain
\begin{equation}\label{eq:etaT1etaT2}
\etab_k^{\curlyT_1} \Perp \etab_k^{\curlyT_2} \given \Zvec^{\curlyF_1}, \Zvec^{\curlyF_2}, \etab_k^\rest, \change{\eelse},
\end{equation}
and therefore $\etab_k^{\curlyT_1}$ and $\etab_k^{\curlyT_2}$ can be updated in parallel \citep{Wilkinson_2006}. 
The log of the full conditional distributions are identical in form to \eqref{eq:theta0cond} with the same extra term mentioned in Sect.~\ref{sec:scale1}, which is simple to compute. Therefore, each of these updates again requires only a small chunk of data, matrices, and subsets of samples of the other process coefficients, so that each of these updates can also be done via a distributed-data architecture. \change{The computational complexity of the operations required to sample from $\etab_k^{\curlyT_1}$ and $\etab_k^{\curlyT_2}$ are approximately $O(|\curlyT_1^{3/2}|)$ and $O(|\curlyT_2^{3/2}|)$, respectively.}

\begin{figure}[t!]
	\begin{center}
		\includegraphics[width=\ifSTCO 0.45\else 0.6\fi\textwidth]{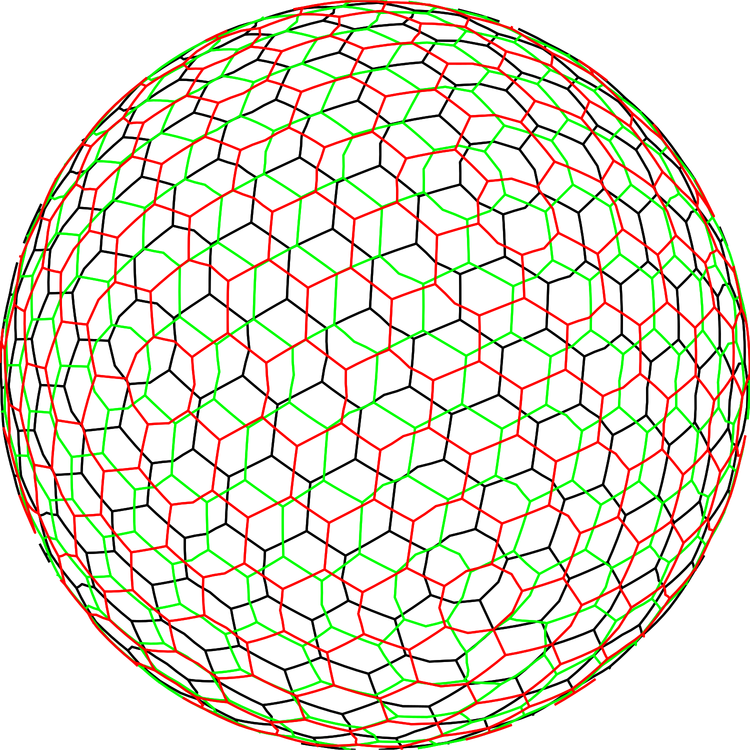} \vspace{0.1in}
		\caption{Example of three tilings on $D = \mathbb{S}^2$  that ensure that basis functions (not shown) are not in the vicinity of a boundary in at least one of the tiles.\label{fig:threeparts} }
	\end{center}
\end{figure}

Even though this sampling \change{strategy re-updates} $\etab_k$, this coloured-tile scheme has got its own boundaries close to which the process coefficients will experience poor mixing. We address this by using three tilings for each $k$, one after another in the Gibbs sampler.  Each tiling is shifted by about a third of a tile, relative to the one before.  This ensures that no element of $\etab_k$ is always near a boundary when Gibbs sampling. We show three such tilings on $\mathbb{S}^2$ in Fig.~\ref{fig:threeparts}. The resulting Gibbs sampler is an instance of a blocked sampler where the blocks are not disjoint; see, for example \citet{Jensen_1995}. \change{Appendix \ref{app:sim_blocking} illustrates the benefit of an alternating tiling scheme on a simple model.}
 
\subsection{Updating the measurement-error variance parameters}\label{sec:epsilon_updates}

Consider now the parameters $\thetab_\epsilon$. The submodel for the Gibbs update of these parameters is made up of
\begin{equation}
  \begin{aligned}
    \Zvec \given \etab, \thetab_\epsilon & \sim \Gau \big(\Amat\etab, \change{\Qmat_\epsilon^{-1}} \big), \\
  \end{aligned}
\end{equation}
\noindent and \change{the prior distribution} \eqref{eq:thetamodel1} with $X = \epsilon$, \change{where  $\Qmat_\epsilon^{-1}$ is diagonal with non-zero elements $[\exp(\bvec_\epsilon(\svec_l)^\top \thetab_\epsilon)]^2, l = 1,\dots,m$.}  As with $\etab_k$ and $\thetab_k$, elements of $\thetab_\epsilon$ are spatially-indexed by the maxima of the basis functions they weight. Let $\curlyI$ contain the index of one element of $\thetab_\epsilon$. The data footprint  $\curlyF$ of $\thetab_\epsilon^\curlyI$ contains the indices of the data points that lie in  $\supp(\bvec_\epsilon^{\curlyI})$.
For two elements of $\thetab_\epsilon$ with indices $\curlyI_1$ and $\curlyI_2$, respectively, and for which $\curlyF_1 \cap \curlyF_2 = \emptyset$, the relevant part of the conditional independence graph is
\begin{equation*}\label{eq:DAGeps}
\hspace{0.7in}  \xymatrix{
  \Zvec^{\curlyF_1} &  \Zvec^\rest & \Zvec^{\curlyF_2} \\
  \thetab_\epsilon^{\curlyI_1} \ar[u] & \thetab_\epsilon^\rest \ar[u] \ar[ul] \ar[ur] & \thetab_\epsilon^{\curlyI_2} \ar[u]    \\
  }
\end{equation*}
because the components of $\Zvec$ are independent given $\thetab_\epsilon$ and everything else, 
and the components of $\thetab_\epsilon$ are IID in the prior.

As with $\etab_1$, the condition $\curlyF_1 \cap \curlyF_2 = \emptyset$ is essential, but this is easily satisfied in our context since $\bvec_\epsilon(\cdot)$ are tent basis functions and the data are point-referenced. In particular, this condition is satisfied whenever $\thetab_\epsilon^{\curlyI_1}$ and $\thetab_\epsilon^{\curlyI_2}$ are not neighbours in the graph implied by $T_\epsilon$. In this case, the full conditional dependence graph of $\thetab_\epsilon$ is just an edge-preserving bijection of the spatial graph $T_\epsilon$, with $\thetab_\epsilon$ as the vertices. \change{We therefore also colour it using four colours, and we update parameters of the same colour in parallel.}  The full conditional distribution,
\begin{equation}
  \pr(\thetab_\epsilon^{\curlyI_1} \given \eelse) \propto \pr(\Zvec^{\curlyF_1} \given \thetab_\epsilon^{\curlyI_1}, \thetab_\epsilon^\rest, \etab) \pr(\thetab_\epsilon^{\curlyI_1}),
\end{equation}
does not have a recognisable distribution,  and so this update uses a Metropolis--Hastings step. \change{Since the measurement errors are independent, computing the accept-reject ratio is computationally inexpensive.}

\subsection{Summary of Gibbs sampler}\label{sec:Gibbssummary}

Algorithm~\ref{alg:spatial} gives a summary of all the stages in one complete pass through the Gibbs sampler. Each sample is generated conditional on $\Zvec$ and the most recent samples of the other random variables, again denoted here as `$\eelse$' for `everything else.'  
When the conditional distribution to sample from is not available in closed form, the update is done via an accept-reject step; in our application in Sect.~\ref{sec:application} we use adaptive random walk Metropolis proposals. In Algorithm~\ref{alg:spatial} we use the superscript $*$ to denote intermediate quantities: these quantities are discarded in the final output of the Gibbs sampler.

\begin{algorithm-float}[t!]
\begin{framed}
\begin{algorithm}
\label{alg:spatial} \textbf{Parallel Gibbs Sampler for the Multi-Scale Process Model}

\begin{enumerate}
\itemsep0em
  \item Sample $\{\etab_0, \thetab_0\} \given \eelse$.

  \item For $k = 1,\dots,K$
    \begin{enumerate}
      \item Sample $\{\etab_k^*, \thetab_k\} \given \eelse$ in parallel ($\ge$ 4 colours). \label{it:thetab1}
      \item  Sample $\etab_k \given \eelse$ in parallel (4 colours). \label{it:etab1}
    \end{enumerate}

  \item Sample $\thetab_\epsilon \given \eelse$ in parallel (4 colours).
  \item Shift the tiling for each $\etab_k, k \ge 1,$ to the next position and go to step 1.
\end{enumerate}

\end{algorithm}
\end{framed}
\end{algorithm-float}

Steps 1--3 in the Gibbs sampler use distributed data, while steps 2 and 3 also use parallel computation to obtain a complete sample in a fixed number of sweeps (usually 4). It is worth re-iterating that, despite the re-updating of $\etab_k$ in step 2(b), updating $\etab_k$ for each colour in step 2(a) is still required in order to ensure the correct stationary distribution is  targeted when going through the colours in sequence \citep[][]{vanDyk_2008}\change{; see Appendix~\ref{app:target}}.

\section{\change{Implementation guidelines}}\label{sec:guidelines}

\change{The favourable computational properties of the updates in Sect.~\ref{sec:inference} can only be taken advantage of if sensible triangulations are used for the process and parameter decompositions. Further, the separation of $Y(\cdot)$ into separate scales introduces identifiability issues, akin to those seen in problems of spatial source separation \citep{Nordhausen_2015}. Therefore, judicious construction of the triangulations and prior distributions on the \change{nominal ranges} (that help to  separate the scales in spectral space) are required. Our informal guidelines in this section are based on point-referenced measurements that have relatively high signal-to-noise ratio; these guidelines may need to be slightly adjusted when this is not the case.}

\subsection{\change{Triangulations}}\label{sec:triangulations}

\paragraph{\change{Process:}} \change{Each process $Y_k(\cdot)$ is associated with a triangulation $T_{\eta_k}$. Recall that, as $k$ increases, the process scale decreases, and therefore we seek triangulations that become finer with increasing $k$. Constructing appropriate triangulations on which to establish the multi-scale process is important. In Appendix \ref{app:sim_coarseness} we empirically show the possible consequences of an inappropriate discretisation scheme for a two-scale process in one-dimensional space.}

\change{A key design criterion when constructing these triangulations is the size of largest matrix that can be efficiently factorised on a single computing node. Using hardware and linear algebra libraries current as of the year 2020, sparse precision matrices of dimension 50000 constructed from a second-order GMRF on a two-dimensional manifold (following appropriate fill-in reducing permutations) can be factorised extremely quickly \change{(in about one second)} on a single computing node. Since in our model $\etab_0$ is updated as a whole (see \eqref{eq:joint0}), a  regular triangulation $T_{\eta_0}$ on $D$ should therefore be constructed such that $r_{\eta_0}$ is between $10^4$ and $10^5$.}

\change{At the other end of the spectrum, the finest process triangulation, $T_{\eta_K}$, should be fine enough that the size of the data footprint associated with each element of $\etab_K$ is small, and preferably in the single digits. This guideline ensures that our multi-scale process model is able to resolve the highest frequency components that may be extracted from the data. If we do not do this, then our predictions at the finest scale are possibly over-smoothed (i.e., the model is underfitting).}

\change{The number of scales $K$ to use is a modelling choice which can be guided by spectral considerations. A good rule of thumb is to ensure that each scale is associated with signals with nominal ranges that approximately span an order of magnitude, and that the triangulations can adequately reproduce signals with these ranges. On a two-dimensional manifold, this rule of thumb implies that $|\etab_{k}| \approx 100|\etab_{k-1}|$, for $k = 1,\dots,K$. In Sect.~\ref{sec:application}, we use two scales with $r_{\eta_0}$ on the order of $10^4$ and $r_{\eta_1}$ on the order of $10^6$.}

\paragraph{\change{Process parameters:}} \change{The update of $\thetab_k^\curlyI$ in \eqref{eq:thetaapprox} requires factorising a square matrix with number of columns equal to the size of its effective process footprint,  $|\curlyT|$, which thus needs to be on the order of $r_{\eta_0}$. Therefore, since $T_{\eta_k}$ becomes finer with increasing $k$, this necessarily means that $T_{\theta_k}$ also needs to become finer with increasing $k$. Specifically, the number of basis functions in $\avec_k(\cdot)$ that have support intersecting $\bvec_k^\curlyI(\cdot)$, for each $\curlyI$, needs to be on the order of $r_{\eta_0}$. This successive refining of $T_{\theta_k}$ with $k$ yields an attractive model where the degree of nonstationarity of the process increases with  $k$. }

\paragraph{\change{Measurement-error standard deviation:}} \change{The coarseness of the triangulation $T_\epsilon$ reflects a prior belief on how quickly (spatially) the measurement-error variance changes; this is application-specific.}

\subsection{\change{Parameter prior distributions}} 

\paragraph{\change{Nominal ranges:}} \change{In Sect.~\ref{sec:triangulations} we associated $T_{\eta_k}$ with nominal ranges spanning approximately an order of magnitude, and this choice can be reflected in our prior distribution for the nominal ranges.  We suggest that $\omega_{\rho_0}$ and $\lambda_{\rho_0}$ are set such that a low percentile of $\pr(\rho_0)$, the 2.5 percentile say, is no smaller than one to two times the average edge length in $T_{\eta_0}$. Similarly, we suggest that $\omega_{\rho_k}$ and $\lambda_{\rho_k}$ are set such that a low percentile of $\pr(\exp(\thetab_{\rho_k}^\curlyI))$, for each $\curlyI$, is no smaller than one to two times the average edge length in $T_{\eta_k}$. This  choice is motivated by the Nyquist--Shannon criterion, which is also used for basis function placement by \citet{Zammit_2012}. From a practical point of view, it encapsulates the fact that signal components with finer scales cannot be captured by this decomposition of $Y_k(\cdot)$ on $T_{\eta_k}$. The upper 2.5 percentile of $\pr(\rho_0)$, corresponding to the highest plausible nominal range of the multi-scale process, can usually be set based on application considerations. The upper 2.5 percentile of $\pr(\exp(\thetab_{\rho_k}^\curlyI)), k = 1,2,\dots,K,$ for all $\curlyI$, can be set to be close to, but higher than, the lower 2.5 percentie of $\pr(\exp(\thetab_{\rho_{k-1}}^\curlyI))$. This ensures some spectral overlap across the scales.}

\paragraph{\change{Process standard deviations:}} \change{The prior distribution $\pr(\sigma_0)$ can be uninformative or based on results from an exploratory data analysis. Since in this model source separation is done via the prior distributions of $\rho_k$, the hyperparameters $\omega_{\sigma_k}$ and $\lambda_{\sigma_k}$ can be set such that $\pr(\exp(\thetab_{\sigma_k}^{\curlyI}))$, for each $\curlyI$, is relatively uninformative or identical to $\pr(\sigma_0)$.} 

\paragraph{\change{Measurement-error standard deviation:}} \change{The hyperparameters $\omega_{\epsilon}$ and $\lambda_\epsilon$ encode the plausible range of values of measurement-error standard deviations, and their choice is therefore application-specific.}

\subsection{\change{Effective process footprint}} \label{sec:guidelines_footprint}

\change{The effective process footprint of each $\thetab_k^\curlyI$ is a design choice. In principle, the footprint needs to be one such that the approximation \eqref{eq:etabkmargapprox} is acceptable. For example, one might use a spatial buffer around $\supp(\bvec_k^\curlyI)$ of width equal to the upper 97.5 percentile of $\pr(\exp(\thetab_{\rho_k}^\curlyI))$ to define the effective process footprint of $\thetab_k^\curlyI$. If this footprint is so large that sampling of $\thetab_k^\curlyI$ becomes prohibitive, then tighter prior distributions (and hence more scales in order to ensure spectral coverage everywhere) would need to be used if this approximation is of concern. Alternatively, one might simulate a number of Markov chains with different computationally inexpensive effective process footprints in parallel, and verify that the target distribution is indeed (approximately) independent of the chosen footprint size. }
\change{As discussed in Sect.~\ref{sec:scale1}, in our application we have taken a minimal approach and used relatively small effective process footprints. This choice does not seem to have had a detrimental effect on the quality of the probabilistic predictions which, as we show next in Sect.~\ref{sec:application}, are considerably better than those obtained from competing models. }

\section{Case study: Spatial modelling and prediction of sea-surface temperature from VIIRS}\label{sec:application}

A sophisticated distributed implementation on a multi-node cluster is required for the sampler to be used with data-set sizes on the order of tens or hundreds of millions and with models with many ($K > 1$) scales. However, a two-scale model with approximately one million basis functions and one million data points can be implemented on a standard multi-core computing node using a straightforward implementation in \texttt{R} \citep{R}, whereby the ``chunks'' required for each parallel sample (implemented via the function \texttt{mclapply}) are loaded from disk when needed. This data size is still considered fairly large in spatial analyses, and we therefore restrict ourselves to this case in this article. An implementation that allows for orders of magnitude more data and basis functions is being developed and will be discussed elsewhere.

We compare inferences from the two-scale variant of the multi-scale spatial process model with those from several other spatial models and approximation methods, specifically a coarse-scale SPDE model; a suite of independent fine-scale SPDE models over spatially contiguous blocks of data; a single-scale model approximated using a full-scale approximation (FSA); a stationary multi-scale process model constructed using a relatively small number of basis functions, and a single-scale process model using the approximations via a nearest-neighbour Gaussian process approximation. These models and approximations were chosen to show that several aspects of the proposed multi-scale model, namely the use of multiple scales, the use of a large number of basis functions, and the ability to model nonstationarity, are crucial to get good predictions in a typical big-data environmental spatial data analysis.

\change{Reproducible code and additional material, such as MCMC trace plots,  are available from \url{https://github.com/andrewzm/SpatialMultiScale}.} 

\subsection{Data}\label{sec:Data}

As data we used global sea-surface temperature (SST) obtained from the Visible Infrared Imaging Radiometer Suite (VIIRS) on board the Suomi National Polar-orbiting Partnership (Suomi NPP) weather satellite on October 14 2014 \citep{Cao_2013}. We sampled one million data points at random from the complete data set \change{of 238 million data points}, intentionally leaving data out from an $8^\circ \times 8^\circ$ box centred at ($155^\circ$W, $0^\circ$N) in the Pacific Ocean, and used these as observation/training data. We then sampled another one million data points at random from the remainder and used these for assessing the quality of our predictions. After discarding data \change{with invalid coordinate entries}, and data points that fell outside our constructed meshes (see Sect.~\ref{sec:model_setup}) we were left with about 900000 data points in each of the training and validation data sets.

\begin{figure*}[t!]
  \includegraphics[width = 0.51\textwidth]{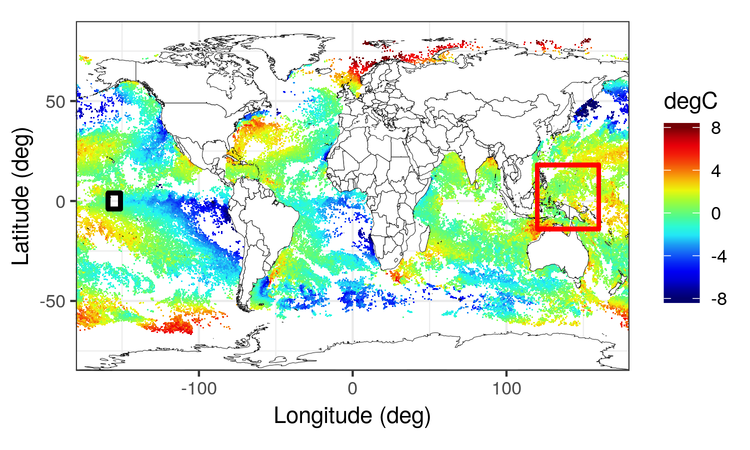}
  \includegraphics[width = 0.48\textwidth]{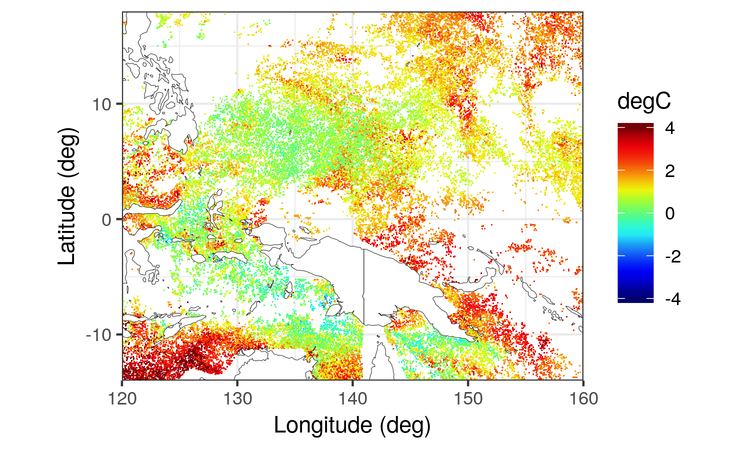}
  \caption{SST residuals used for training the model. Left panel: Global view of residuals, with the $8^\circ\times 8^\circ$ box in the Pacific Ocean marked in black \change{and the box delineating the region depicted in the right panel marked in red}. Right panel: Residuals in an area of the Western Pacific around Papua New Guinea (note the different colour scale).\label{fig:data}}
\end{figure*}

We then detrended the training and validation data sets using a linear model with an intercept, the latitude coordinate, and the square of the latitude coordinate, as covariates. The residuals from the training data are what we call $\Zvec$ in Sect.~\ref{sec:model}, while we denote the residuals from the validation data set as $\Zvec_v$. In Fig.~\ref{fig:data} we show the residuals $\Zvec$ on the globe together with the $8^\circ \times 8^\circ$ box  (left panel) and a zoomed-in view of these residuals around Papua New Guinea (right panel). We can see from the figure that the spatial data are very irregularly-sampled in space, and that the distance from anywhere in the ocean to the nearest data point can range from a few kms to thousands of kms. The data also reveal both large-scale and small-scale features that need to be modelled.

\subsection{Model Setup}\label{sec:model_setup}

\begin{figure}[t!]
  \begin{center}
    \includegraphics[width = \ifSTCO 0.45\else 0.6\fi\textwidth]{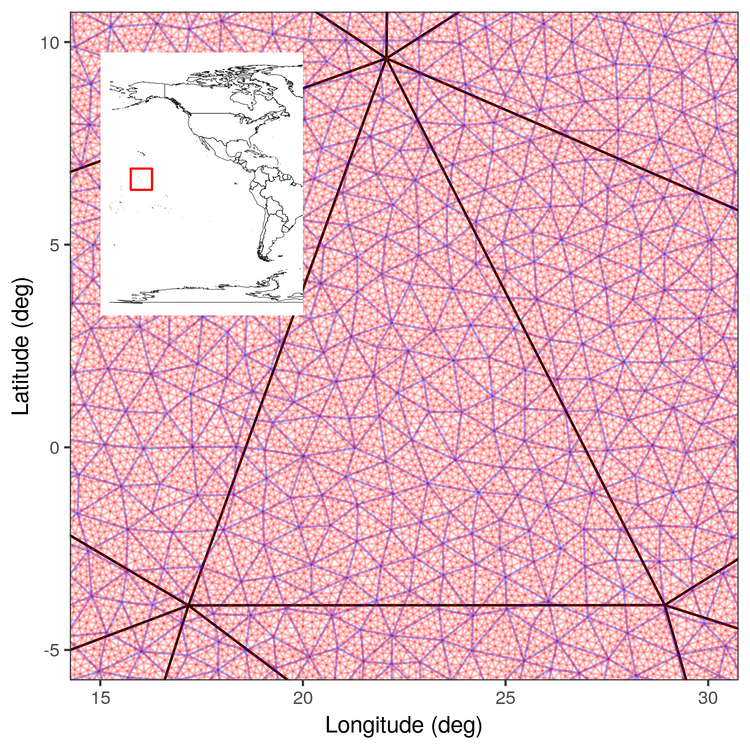}
    \end{center}
  \caption{The triangulations $T_{\eta_1}$ (red), $T_{\eta_0}$ (blue), and $T_{\epsilon}~ (= T_{\theta_1})$ (black) shown here for a small region in the Pacific Ocean as marked by the red rectangle in the inset. \label{fig:meshes}}
\end{figure}

We constructed the Delaunay triangulations on $\mathbb{S}^2$ with a coarse boundary around land masses \cite[used in a study by][]{Simpson_2016} using the \change{\texttt{inla.mesh.create}} function in \texttt{INLA}, and fixed $\nu_0 = \nu_1 = 1$. We determined the approximate number of vertices in the triangulations as follows, using the guidelines in Sect.~\ref{sec:guidelines}. First, since it is required to factorise $\Qmat_0$ we made sure that $T_{\eta_0}$ contains only a few tens of thousands of vertices, in our case it contains $r_{\eta_0} = 38274$ vertices. Second, since $Y_1(\cdot)$ models the small-scale variability and we are only considering two scales, we require $T_{\eta_1}$ to be spatially dense with respect to the observed data. In our case we constructed a triangulation $T_{\eta_1}$ with 942349 vertices. Using this triangulation, the proportion of elements in $\etab_1$ which have a data footprint size in the single digits is 92\%.  Finally, we want the nonstationarity in the fine-scale process to vary  smoothly in space, and we therefore let $T_\epsilon = T_{\theta_1}$ have only a few hundred vertices, in our case 205 vertices. Using this triangulation the largest effective process footprint is of size 27198, which is on the same order as $r_{\eta_0}$ as desired. The triangulations on a small region of the Pacific Ocean are shown in Fig.~\ref{fig:meshes}.

\begin{figure}[t!]
  \begin{center}
  \includegraphics[width = 0.45\textwidth]{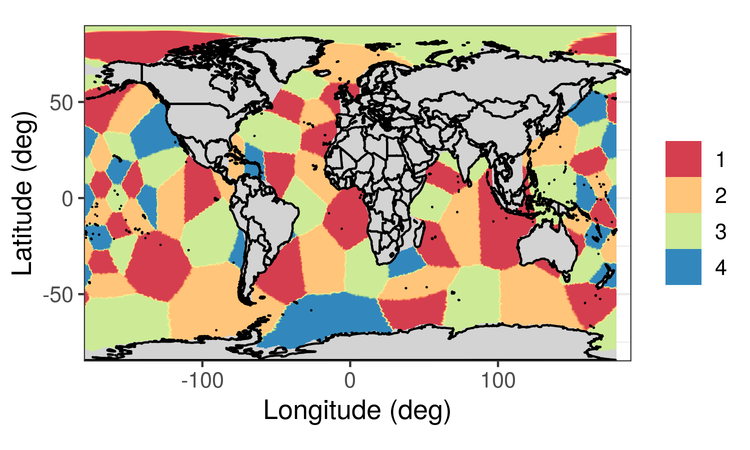}
  \includegraphics[width = 0.45\textwidth]{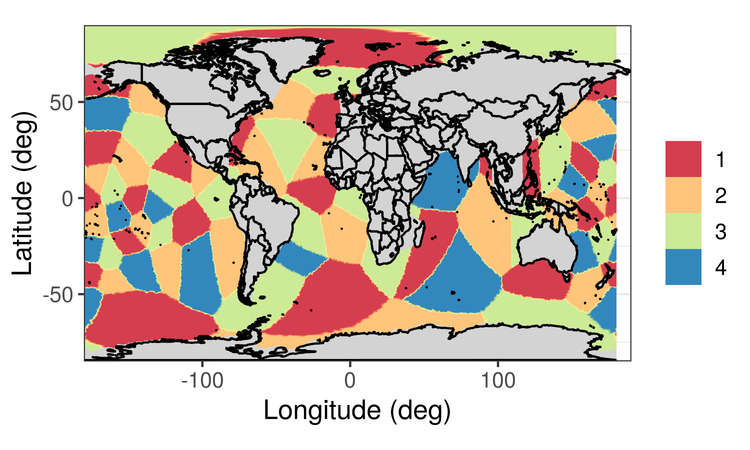}\\
  \includegraphics[width = 0.45\textwidth]{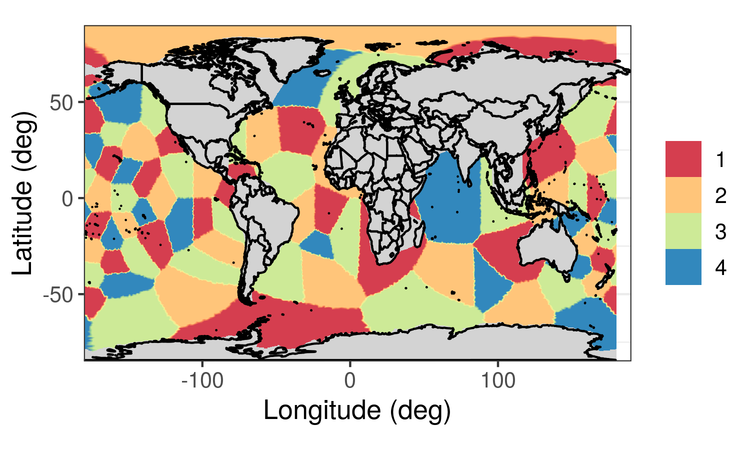}
  \end{center}
  \caption{Colourings of the three tilings of $\etab_1$. \label{fig:colourings}}
\end{figure}

In order to re-sample $\etab_1$ as discussed in Sect.~\ref{sec:weights1}, we constructed three partitionings of $\etab_1$ from three spatial tilings of $D$. The first tiling was done using the third resolution of an  Icosahedral Snyder Equal Area Aperture 3 Hexagonal  (ISEA3H) grid on the surface of the sphere \change{\citep{Sahr_2008}}. The second and third tilings were then done by shifting the ISEA3H grid north and east, respectively; see Fig.~\ref{fig:threeparts}. To improve mixing of the \change{Markov chain}, spatial tiles that contained less than 100 data points or less than 200 basis functions at scale $k = 1$, were  merged with neighbouring tiles. 
The three  tilings and colourings 
are shown in Fig.~\ref{fig:colourings}.

We constructed the prior distributions following the guidelines of Sect.~\ref{sec:guidelines}. As prior distribution for $\rho_0$ we used a lognormal distribution such that $(\rho_{0,0.025}, \rho_{0,0.975}) = (300, 10000)$ km, where $\rho_{0, q}$ denotes the $q$th quantile of the distribution. Our choice of $\rho_{0,0.025}$ stems from us having a maximum edge length of 150 km when constructing $T_{\eta_0}$, while our choice of $\rho_{0,0.975}$ presents a soft maximum on what we believe the spatial correlations in SSTs are (10000 km equates to approximately half the greatest east-west span of the Pacific Ocean). On the other hand, as prior distribution for $\thetab_{\rho_1}^{\{i\}}, i = 1,\dots,r_{\rho_1},$ we used a \change{Gaussian} distribution such that \change{$(\exp(\thetab_{\rho_1,0.025}^{\{i\}}), \exp(\thetab_{\rho_1,0.975}^{\{i\}})) = (30, 900)$ km}. The lower quantile was selected on the basis that we used 30 km as our maximum edge size when constructing $T_{\eta_1}$, while the upper quantile was selected to ensure that there is sufficient spectral overlap between $Y_0(\cdot)$ and $Y_1(\cdot)$. As for the marginal process standard deviation, we used prior distributions such that \change{$(\sigma_{0,0.025}, \sigma_{0,0.975}) = (\exp(\thetab_{\sigma_k,0.025}^{\{i\}}), \exp(\thetab_{\sigma_k,0.975}^{\{i\}})) = (0.1, 6)^\circ$C},  where our upper quantile was selected from the fact that the empirical standard deviation of the SST residuals is approximately equal to 2$^\circ$C. Finally, as prior for $\thetab_{\epsilon}$ we used a Gaussian distribution such that $(\exp(\thetab_{\epsilon,0.025}^{\{i\}}), \exp(\thetab_{\epsilon,0.975}^{\{i\}})) = (0.5, 5)^\circ$C, which reflects our prior belief that it is unlikely that the measurement error standard deviations are less than 0.5$^\circ$C or larger than $5^\circ$C.

\change{As initial values for our parameters in the MCMC algorithm we used samples of $\thetab_0, \thetab_1,$ and $\thetab_\epsilon$ drawn from their respective prior distributions. As initial values for $\etab_0$ we used least-squares estimates from the model $\Zvec = \Amat_0\etab_0 + \epsilonb$, perturbed using independent samples from a Gaussian distribution with mean zero and variance 0.04. As initial values for $\etab_1$ we used independent samples from a Gaussian distribution with mean zero and variance 0.01. Our adaptive random-walk Metropolis proposals were initialised to have variance 0.001; adaptation was done every 30 iterations for the first 2000 iterations.}

The Gibbs sampler algorithm in Sect.~\ref{sec:Gibbssummary} was run for 10000 iterations. The first 5000 samples were discarded as burn-in and the remaining 5000 were thinned by a factor of 50, to yield 100 samples from the posterior distribution over all the states and parameters. It took approximately 5 days of computing time to obtain 10000 samples using our implementation.

\begin{figure}[t!]
  \begin{center}
  \includegraphics[width = \ifSTCO 0.45\textwidth \else 0.8\textwidth \fi]{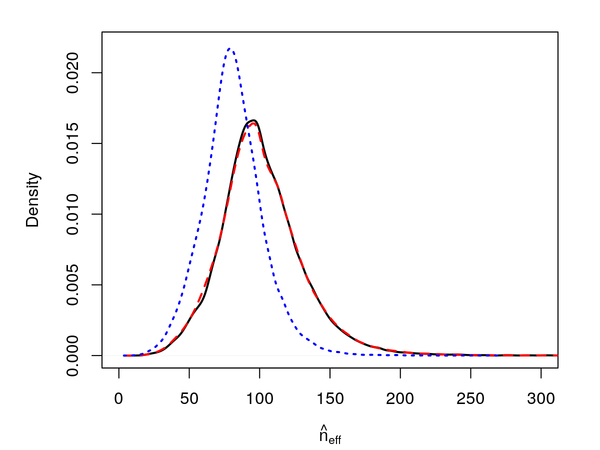}
  \end{center}
  \caption{\change{Empirical density of the estimated effective sample size $\hat{n}_{\eff}$ from a chain with independent samples from a normal distribution (black, solid line), a chain where each trace is a first-order autoregressive process with auto-correlation parameter 0.1 (blue, dotted line), and the chain's traces for $Y(\cdot)$ evaluated at approximately 350000 locations using our sampler (red, dashed line).} \label{fig:neff}}
\end{figure}

\change{The long run-time of our algorithm made a detailed test of MCMC convergence impractical.  Instead, we made an
	informal test of convergence by analysing samples of the process $Y(\cdot)$  evaluated at approximately 350000 points regularly interspersed throughout the ocean. Specifically, we computed the effective sample sizes \citep[see Eq.~11.8 of][]{Gelman_2013} of the thinned chain at each of these locations, and compared the empirical density of these sizes to that constructed from independent samples from a Gaussian distribution, and to that constructed from traces that are first-order autoregressive processes with auto-correlation coefficient 0.1. In Fig.~\ref{fig:neff} we see that the empirical density is very sensitive to auto-correlation, and that the effective sample sizes from our sampler are virtually indistinguishable from those computed from independent samples. This indicates that most likely we do indeed have independent samples from the posterior distribution. Visual inspection of a random selection of trace plots also suggested the absence of auto-correlation.} 
	
	\change{Traces of other model components were not all as uncorrelated: While, for example, traces for $\etab_0$,  $\etab_1$, and $\thetab_\epsilon$, had median effective sample sizes of 87, 90, and 90, respectively, those for $\thetab_{\sigma_1}$ and $\thetab_{\rho_1}$ had median effective sample sizes of 52 and 46, respectively.  While the latter low effective sample sizes imply that we should be cautious when making inferential statements on the nominal ranges and variances, they do not necessarily suggest that our  \emph{process} predictions could benefit from a longer run. This is a consequence of weak identifiability of the process model we are using; see \citet{Besag_1995, Everly_2000} and \citet{Gelfand_2001} for more discussion. Moreover, the purpose of our approach is to predict, and for prediction the ultimate goal is to maximise out-of-sample performance.  In this respect the key finding is that our approach is well-calibrated out-of-sample, and that it considerably outperforms other approaches, as we describe in Sect.~\ref{sec:results}.}

\subsection{Other models}

We compared the predictions of the proposed model (SPDE2) to several other models and approximations commonly used in applications containing large spatial data sets. We describe these in more detail below.

\paragraph{Global coarse-scale model (SPDE0):} We used \texttt{INLA} to fit the SPDE \citep{Lindgren_2011} in \eqref{eq:SPDE0} to the entire data set via a GMRF on the coefficients of the tent basis functions on $T_{\eta_0}$. We let $\nu = 1$, and used the same prior distributions for $\rho_0$ and $\sigma_{0}$ as we did for our two-scale spatial model in Sect.~\ref{sec:model_setup}. We compare to this model to demonstrate the benefit of using a second, small-scale process when modelling such large spatial data sets. It took approximately 30 minutes of computing time to fit the model and obtain predictions with SPDE0. \change{The nominal range for this model was estimated to be approximately 800 km}.

\paragraph{Multiple spatially-independent small-scale processes (SPDE1-indep):} We used \texttt{INLA} to independently fit SPDEs with spatially-invariant parameters to data that lie in each of the tiles shown in Fig.~\ref{fig:colourings}, top panel. For each set of tile indices $\curlyT$ in our model SPDE2, we generated a fine-resolution mesh that has a similar number of basis functions as $|\curlyT|$, and a few more around the boundary of the partition to reduce boundary effects. We let $\nu = 1$, and used the same prior distributions for (the now spatially-invariant) $\sigma$ and $\rho$ in each tile as we did for our two-scale process model in Sect.~\ref{sec:model_setup}. These independent models were used to predict the noisy process at validation data locations that lie within their associated tile. Results from this model are used to demonstrate the benefit of having a large-scale process that can borrow strength across large spatial scales when modelling large spatial data. It took approximately 6 hours to sequentially fit the models and obtain predictions with SPDE1.

\paragraph{Multi-scale stationary process with a relatively small number of basis functions via LatticeKrig (LTK):} We used \texttt{LatticeKrig} \citep{Nychka_2015} to fit a 3-resolution LatticeKrig \change{model on the cylinder} 
comprising a total of 137577 basis functions. Note that many of these basis functions (approximately one third) are over land. We fitted three LTK models with the parameter \texttt{a.wght}, which dictates the spatial range, fixed to 4.01, 5.01, and 6.01, respectively. In Sect.~\ref{sec:results} we only show results for the case \texttt{a.wght} = 6.01, which provided slightly better predictions than the other two cases (all cases gave very similar predictions). LTK models are generally limited to $r = 100000$ to $200000$ basis functions since they require factorisation of precision matrices of dimension $r\times r$ when being fit. Results from this model are used to demonstrate that hundreds of thousands of basis functions are still likely to be insufficient when modelling large spatial data, despite the use of multiple scales. Fitting and predicting (via 30 conditional simulations) with LTK required approximately one day of computing time.

\paragraph{Single-scale process approximated using a full-scale approximation (FSA):} The full-scale approximation \citep{Sang_2012} to a Mat{\'e}rn covariance function with smoothness $\nu = 1$ was implemented on $\mathbb{R}^2$ using 100 knots randomly placed on the surface of the sphere before projection onto the plane. The number of blocks on which to approximate the residual field was set to 9000 and data were attributed to each of these blocks using a k-means clustering algorithm on the lon--lat coordinates of the data. This choice ensured that the number of data points in each block was computationally feasible at about 100 data points per block. It took approximately two days to fit and predict using the FSA. \change{The nominal range for this model was estimated to be approximately 850 km}.

\paragraph{Single-scale process approximated using nearest neighbours (NNGP): } We used the \texttt{R} package \texttt{spNNGP} to implement a conjugate version of the nearest-neighbour Gaussian process \citep[NNGP,][]{Finley_2019}, where NNGPs are fitted for several plausible values of the latent-process range and variance, and cross-validation is used to select the best of these models for prediction. In each of these models an inverse-gamma prior distribution is used for the observation measurement-error variance, which yields closed-form predictive distributions that are quick to evaluate. In our implementation we set the number of neighbours to 15, the covariance function to a Mat{\'e}rn covariance function with $\nu = 1$, and used a fine grid for the latent-process range and variance, ensuring that an interior point of this grid was selected as optimal. It took approximately three hours to find the optimal NNGP model through cross-validation and to predict using the optimal model. \change{The nominal range for this model was estimated to be approximately 700 km}. 

\vspace{0.2in}

Results from the FSA and NNGP are used to show that single-scale process models, which use local methods when predicting, are not competitive against models like SPDE2 which are able to borrow strength across large spatial scales when predicting over large gaps.

\subsection{Results}\label{sec:results}

\begin{figure*}[t!]
  \includegraphics[width = 0.49\textwidth]{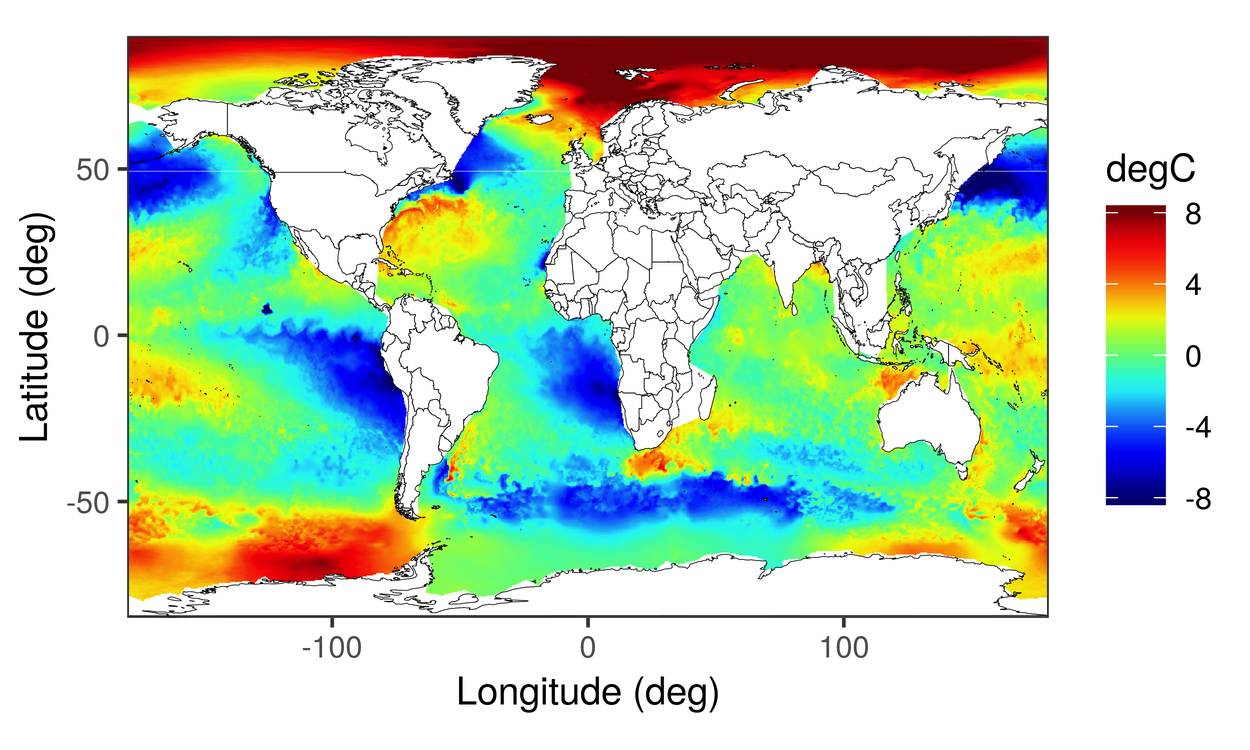}
  \includegraphics[width = 0.49\textwidth]{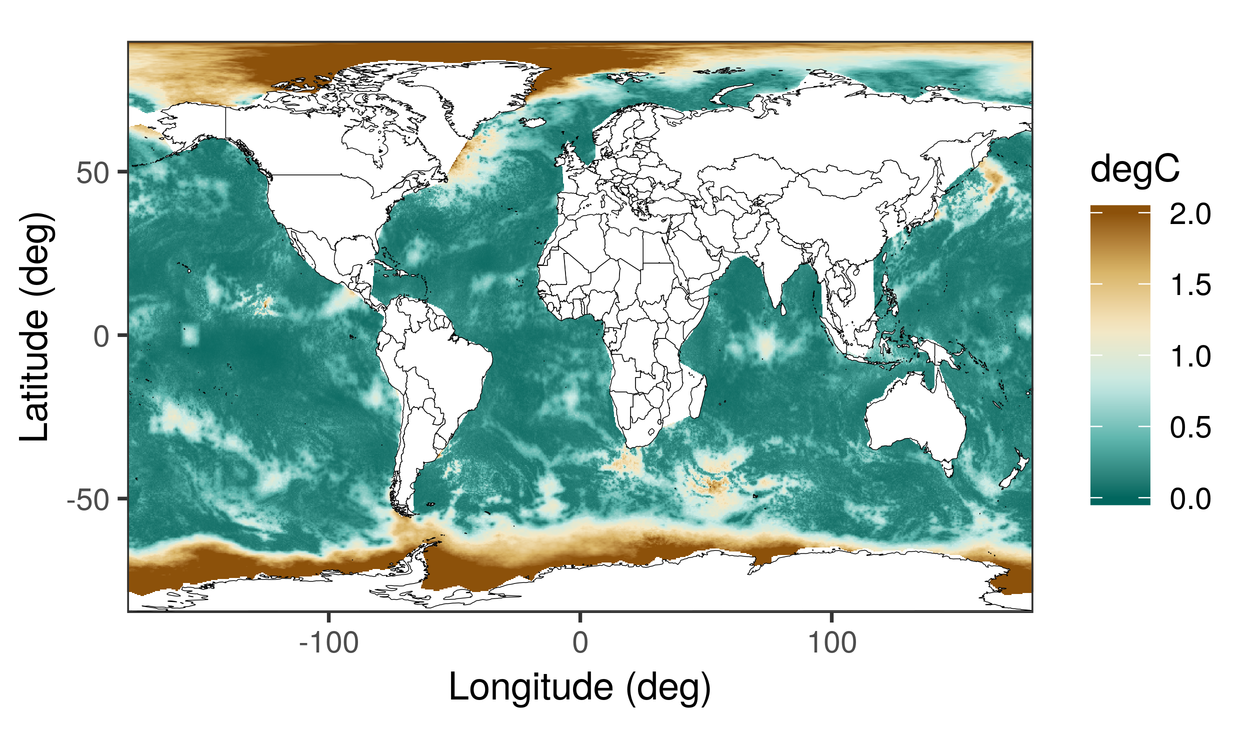}
    \includegraphics[width = 0.49\textwidth]{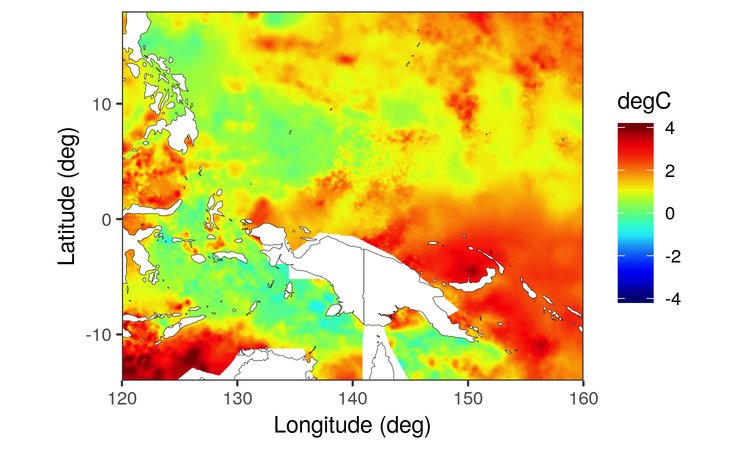}
  \includegraphics[width = 0.49\textwidth]{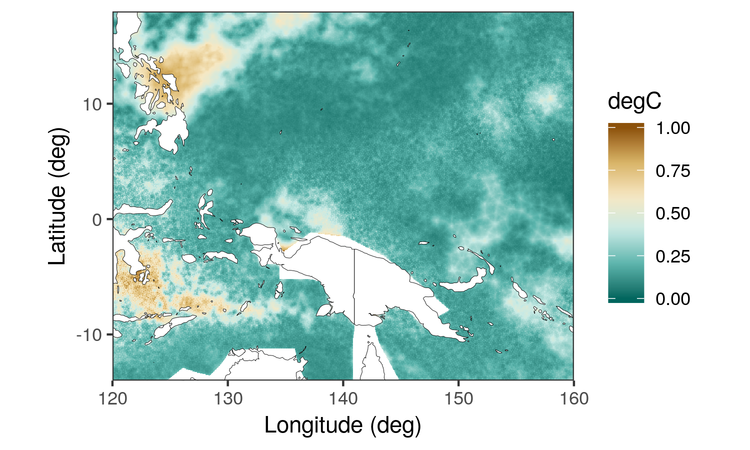}  
  \caption{Global predictions (posterior means, top-left panel) and prediction standard errors (posterior standard errors, top-right panel) of $Y(\cdot)$ from the two-scale spatial-process model (SPDE2) when fitted to the data shown in Fig.~\ref{fig:data}, left panel. Predictions and prediction standard errors corresponding to the region depicted in Fig.~\ref{fig:data}, right panel, are shown in the bottom-left and bottom-right panels, respectively.\label{fig:Ypred}}
\end{figure*}

In Fig.~\ref{fig:Ypred} we show predictions (posterior means) and prediction standard errors (posterior standard errors) for $Y(\cdot)$ using the two-scale spatial-process model SPDE2. The detail in these maps is very high -- the one million basis functions effectively model the SST at a 30 $\times$ 30 km resolution globally. \change{One should be cautious when interpreting the spatially-varying parameters (see Sect.~\ref{sec:model_setup}), however it is evident that t}he posterior expectations of $\thetab_{\epsilon}, \thetab_{\sigma_1}$, and $\thetab_{\rho_1}$ \change{vary} considerably. In particular, \change{the elements in $\E(\exp(\thetab_{\sigma_1}) \mid \Zvec)$ had an interquartile range of $(0.22, 0.85)^\circ$C while those in $\E(\exp(\thetab_{\rho_1}) \mid \Zvec)$ had an interquartile range of $(40, 220)$ km. This suggests that there is ample nonstationarity at small scales that is being captured by SPDE2. These posterior expectations can also be used to interpret the behaviour of the process at each scale. For example, the posterior expectations of the \change{nominal ranges} at $k = 1$ (ranging from a few dozen to a few hundreds of kms) are smaller than $\E(\rho_0 \mid \Zvec) \approx 1700$ km, as expected due to the imposed prior distributions and triangulations. Also, $\E(\sigma_0 \mid \Zvec) = 1.84^\circ$C, and therefore there is more power being allocated to the signal at the coarse scale than at the fine scale. Again, this was expected since power spectral densities tend toward zero at large frequencies. More intuition into the behaviour of the process at the separate scales can be obtained by visualising predictions of $Y_0(\cdot)$ and $Y_1(\cdot)$ separately; we show one such plot centred on the Brazil--Malvinas confluence in our online code repository.}

We compare the predictions from all the models in terms of the root-mean-squared prediction error (RMSPE), continuous-ranked probability score (CRPS), 90\% interval score (IS90), and the 90\% coverage (Cov90) at the validation-data locations \change{\citep{Gneiting_2007}}. To assess the models' ability to predict and quantify uncertainty correctly at validation-data locations that are in regions of both dense and sparse training data, we split the validation data set into three: $\Zvec_v^{(1)}$ contains all the validation data outside the $8^\circ\times 8^\circ$ box in the Pacific Ocean (see Sect.~\ref{sec:Data}) that are in the vicinity of training data (specifically, a $1.5^\circ \times 1.5^\circ$ box centred on each datum in $\Zvec_v^{(1)}$ contains at least one training datum within it); $\Zvec_v^{(2)}$, which contains the validation data not in the vicinity of training data and outside the $8^\circ\times 8^\circ$ box; and $\Zvec_v^{(3)}$ that contains the validation data within the $8^\circ\times 8^\circ$ box. The dimensions of $\Zvec_v^{(1)}, \Zvec_v^{(2)}$, and $\Zvec_v^{(3)}$ were $897282, 293$, and $6964$, respectively.  

\begin{table}[t!]
  \caption{\change{Diagnostic results on the validation data $\Zvec_v^{(1)}$ (validation data in the vicinity of training data), $\Zvec_v^{(2)}$ (validation data not in the vicinity of training data), and $\Zvec_v^{(3)}$ (validation data in the $8^\circ \times 8^\circ$ box).}\label{tab:results_all}}
  \begin{center}
  \resizebox{\ifSTCO\columnwidth\else 0.6\columnwidth\fi}{!}{
 \begin{tabular}{clcccc}
   \hline\noalign{\smallskip}
Data &  Model & RMSPE & CRPS & IS90 & Cov90 \\
   \hline\noalign{\smallskip}
                &   FSA & 0.353 & 0.185 & 1.659 & 0.915 \\
                &   LTK & 0.378 & 0.200 & 1.752 & 0.913 \\
$\Zvec_v^{(1)}$ &   NNGP & {\bf 0.345} & 0.180 & 1.632 & 0.913 \\
                &    SPDE0 & 0.406 & 0.216 & 1.889 & 0.922 \\
                &   SPDE1-indep \hspace{-0.15in} &  0.346 & 0.180 & 1.556 & 0.919 \\
                &   SPDE2 & {\bf 0.345} & {\bf 0.176} & {\bf 1.479} & {\bf 0.907} \\   \hline\noalign{\smallskip}
                &   FSA & 1.001 & 0.540 & 4.489 & 0.928 \\ 
                &   LTK & 0.763 & 0.430 & 4.065 & 0.959 \\ 
$\Zvec_v^{(2)}$ &   NNGP & 0.799 & 0.464 & 4.420 & 0.990 \\ 
                &   SPDE0 & 0.699 & 0.377 & 3.170 & 0.928 \\ 
                &   SPDE1-indep \hspace{-0.15in}  & 0.728 & 0.382 & 3.278 & 0.935  \\
                &   SPDE2 & {\bf 0.680} & {\bf 0.345} & {\bf 2.666} & {\bf 0.894} \\ \hline\noalign{\smallskip}
                &   FSA & 0.460 & 0.325 & 3.570 & 0.993 \\
                &   LTK & 0.463 & 0.339 & 3.748 & 0.986 \\
$\Zvec_v^{(3)}$ &   NNGP & 0.515 & 0.422 & 5.000 & 0.996 \\
                &   SPDE0 & 0.425 & 0.282 & 2.926 & 0.995 \\
                &   SPDE1-indep \hspace{-0.15in} & 0.461 & 0.252 & 1.830 & {\bf 0.916} \\
                &   SPDE2 & {\bf 0.367} & {\bf 0.210} & \bf{1.798} & 0.956 \\ \hline\noalign{\smallskip}
 \end{tabular}}
 \end{center}
\end{table}

Diagnostics of the marginal prediction distributions from each model are provided in Table~\ref{tab:results_all}. \change{In the first six rows of} Table~\ref{tab:results_all} we show diagnostics for when prediction locations are in the vicinity of data points. As expected, the models that rely on a relatively small number of basis functions (SPDE0 and LTK) give relatively poor predictions. On the other hand, the use of a high number of basis functions (SPDE2 and SPDE1-indep) and the single-scale models that \change{use local methods when predicting}  (FSA and NNGP) give better predictions. All methods give the correct coverage. What we deduce from these results is well-known: When doing spatial prediction on large data sets there is little benefit in using a global model when predicting at locations close to observed data \citep[unless the signal-to-noise-ratio is very low; see, e.g., ][]{Zammit_2018} if the target of the spatial analysis is pointwise process prediction. There is also likely to be little benefit to be gained from using nonstationary process models since the inferences are predominantly data-driven and not model-driven. Rather, it is crucial to use a latent process model that is able to reproduce the fine-scale variation in the underlying process.

\begin{figure}[t!]
  \begin{center}
  \includegraphics[width = 0.95\linewidth]{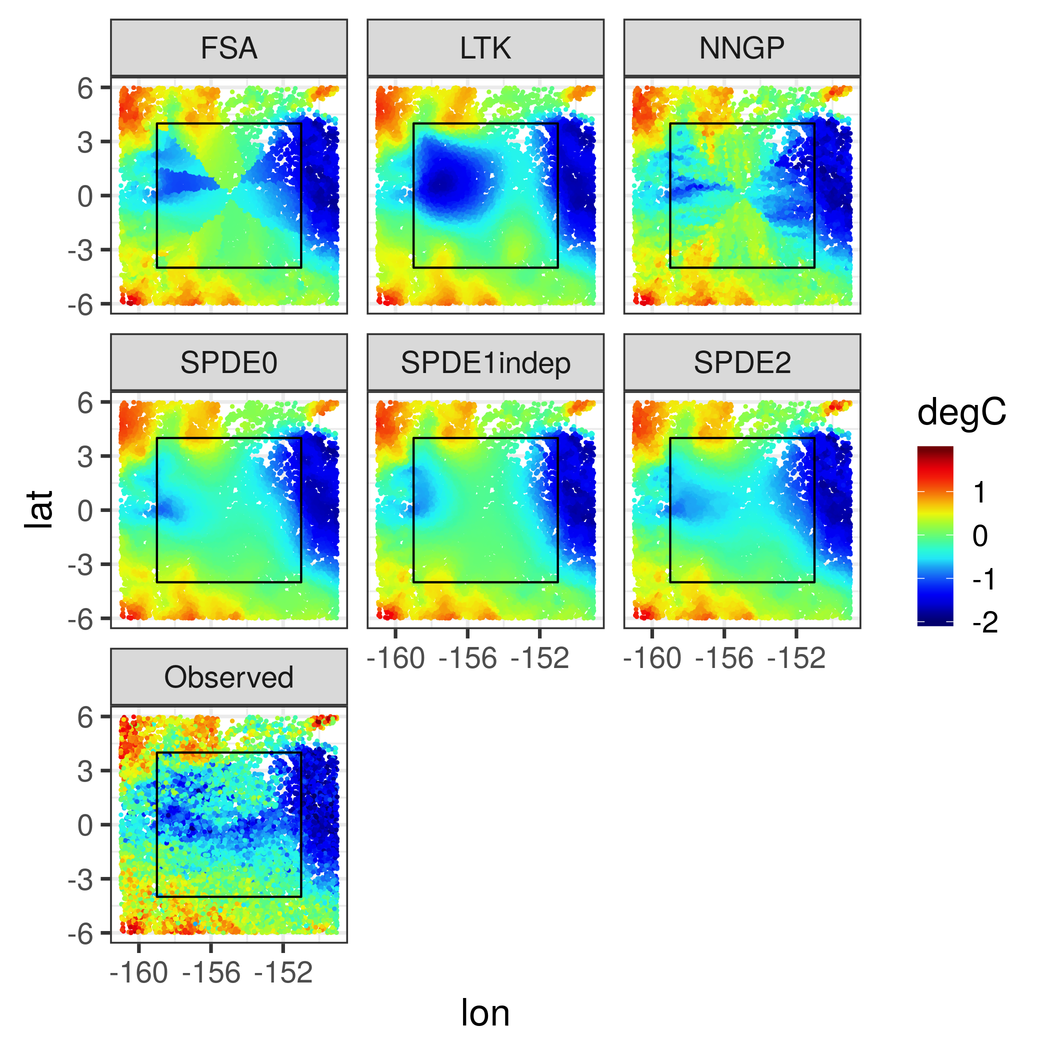}
  \end{center}
  \caption{\change{Process predictions at the validation locations $\Zvec_v^{(3)}$ inside the $8^\circ \times 8^\circ$ region (black box) in which data was left out. The left-out data in this region are shown in the bottom-left panel.} \label{fig:Z3_pred}}
\end{figure}

\change{In the second six rows of Table~\ref{tab:results_all} we show diagnostics for when prediction locations are not in the vicinity (through the $1.5^\circ \times 1.5^\circ$ boxes described earlier) of observations. The models LTK and SPDE0 are now not compromised as much for their use of relatively few basis functions. On the other hand, since FSA and NNGP utilise only a few data points when predicting, they begin to give poorer predictions. SPDE1-indep performs surprisingly well in this scenario, but slightly worse than SPDE0. All methods performed worse than SPDE2. Finally, in the last six rows of Table~\ref{tab:results_all} we show diagnostics for validation data inside the $8^\circ \times 8^\circ$ lon-lat box, while in Fig.~\ref{fig:Z3_pred} we show the predictions at the data locations. We see that predicting using `nearest data'  generates unwanted artefacts in the case of the FSA and the NNGP. We expected LTK to perform well in this case; however, it appears to have over-fit at the region's boundary, resulting in a large negative prediction in the region's interior.  The two-scale model outperforms the others by quite some margin in terms of RMSPE and CRPS since it is able to correctly predict the region of low SST anomaly traversing this box from west to east, while resolving fine scales on the region's boundary. Interestingly all models except SPDE1-indep, and to a lesser extent SPDE2, are largely underconfident in this unobserved area.}

\section{Conclusion}\label{sec:conclusion}

The majority of models and inference techniques developed for analysing massive spatial data sets are not designed to \change{exploit} the multi-scale, nonstationary nature of the underlying process. The result is the widespread use of \change{techniques} that are able to provide good predictions at the fine scale (e.g., NNGP) or at the coarse scale when the number of basis functions for decomposition is capped for computational reasons\change{, but not both}. In this article we have proposed a multi-scale model where the degree of nonstationarity increases with the scale of the component processes, together with an approximate inference algorithm that is scaleable with both data and model size. We have shown that it outperforms other state-of-the-art approaches that are amenable to big data scenarios, but that are unable to capture the complexities of the underlying process due to either the use of a few basis functions, or the use of \change{local methods that only take into account a small subset of the data when predicting}. We conclude that \change{using information at multiple scales is important for accurate prediction} in environmental applications such as the analysis of SST\change{, particularly in regions where remote sensing data is sparse.} \change{We used SPDEs to model the processes at each scale, but other spatial GMRFs (such as conditional autoregressive models) can be used instead. However, we anticipate that constructing appropriate grids, modelling the nonstationary structure, and eliciting interpretable prior distributions for the parameters would become more challenging.}

There are several avenues for future work. First, while the approximation \eqref{eq:etabkmargapprox} is a sensible one under the guidelines of Sect.~\ref{sec:guidelines}, we do not provide any theoretical bound on the effect of the approximation on the target distribution, which is not straightforward to derive. Second, while our implementation in Sect.~\ref{sec:application} used chunked-up data and models when doing inference, these were saved and loaded to disk when sampling via graph colouring. Sophisticated implementations include one where static objects (e.g., data and basis-function matrices) reside permanently in memory on dedicated nodes \citep{Katzfuss_2017b}, while communication of selected samples between the nodes is done efficiently via message passing (e.g., using MPI). Such an implementation would speed up the Gibbs sampler, and is necessary for the consideration of higher scales and data sets that are one to two orders of magnitude larger than that considered in Sect.~\ref{sec:application}. Third, there is considerable interest in making inference with large \emph{non-Gaussian} spatial data. Parallelisation is still possible for this case, but our computational framework would need to be modified slightly since the process coefficients cannot be sampled directly when the data is non-Gaussian. Specifically, when carrying out an update corresponding to a colour, the process coefficients would have to be proposed and then accepted or rejected jointly with the parameter coefficients; see, for example, \cite{Knorr_2002}.

\ifSTCO
\begin{acknowledgements}
\else
\section*{Acknowledgements}
\fi

We thank  Yuliya Marchetti for providing the sea-surface temperature data set, Bohai Zhang for providing the MATLAB code for the implementation of the FSA, Matt Moores for general discussions on improving MCMC mixing, and Quan Vu for providing comments on an early version of this manuscript. AZ--M was supported by the Australian Research Council (ARC) Discovery Early Career Research Award, DE180100203.

\ifSTCO
\end{acknowledgements}
\fi

\ifSTCO
\bibliographystyle{spbasic}      
\else
\bibliographystyle{apalike}      
\fi
\bibliography{GCbib}

%
%

\appendix
\renewcommand{\theequation}{\thesection.\arabic{equation}}

\ifSTCO
\color{blue}
\fi

\section{\change{Targeted distribution of the Markov chain}}\label{app:target}

In Sect.~\ref{sec:scale1} we asserted that it is important to update $\etab_k$ concurrently with $\thetab_k$, even though $\etab_k$ is later re-updated (Sect.~\ref{sec:weights1}). If one does not do this, $\etab_k^{\rest}$ in \eqref{eq:etaT1etaT2} would be `out of sync' with the updated parameters $\thetab_k$; as a consequence, when $\etab_k^{\curlyT_1}$ or $\etab_k^{\curlyT_2}$ is updated in \eqref{eq:etaT1etaT2}, an incorrect distribution would be targeted. This phenomenon occurs when marginalising \citep[termed `trimming' by ][]{vanDyk_2008} in Gibbs samplers.

We show the importance of resampling on a very simple spatial model, where we have two sets of parameters, $\thetab_1$ and $\thetab_2$, and two sets of basis-function coefficients, $\etab_1$ and $\etab_2$. In what follows we omit conditioning on the data $\Zvec$, since this is implicit in all the distributions. We denote the target (posterior) distribution as $\pr_0(\Thetab)$ where $\Thetab \equiv \{\thetab_1, \thetab_2, \etab_1, \etab_2\}$. In MCMC we seek a transition kernel $q(\Thetab' \mid \Thetab)$ such that
\begin{equation}
\pr_1(\Thetab') \equiv \int \pr_0(\Thetab)q(\Thetab' \mid \Thetab)\intd \Thetab = \pr_0(\Thetab'). \label{eq:pr1}
\end{equation}
If \eqref{eq:pr1} holds, then we say that the Markov chain preserves the target distribution, $\pr_0$.

In a vanilla Gibbs sampler, one constructs the transition kernel from full conditional distributions of the target distribution:
\begin{align}
  q(\Thetab' \mid \Thetab) = & \pr_0(\thetab_1' \mid \thetab_2, \etab_1, \etab_2)\pr_0(\thetab_2' \mid \thetab_1', \etab_1, \etab_2) \times \nonumber \\
  & \pr_0(\etab_1' \mid \thetab_1', \thetab_2', \etab_2)\pr_0(\etab_2' \mid \thetab_1', \thetab_2', \etab_1').  \label{eq:q1}
 \end{align}
Successive updating of the parameters in this fashion preserves the target distribution. To see this, substitute \eqref{eq:q1} in \eqref{eq:pr1} to obtain
\begin{align*}
  \pr_1(\Thetab')  & = \iiiint \pr_0(\thetab_1, \thetab_2, \etab_1, \etab_2)\pr_0(\thetab_1' \mid \thetab_2, \etab_1, \etab_2) \nonumber \\
  & ~~~\times \pr_0(\thetab_2' \mid \thetab_1', \etab_1, \etab_2)\pr_0(\etab_1' \mid \thetab_1', \thetab_2', \etab_2) \\
  & ~~~\times \pr_0(\etab_2' \mid \thetab_1', \thetab_2', \etab_1')\intd \thetab_1\intd \thetab_2 \intd \etab_1 \intd \etab_2 \\
  & = \iiint \pr_0(\thetab_2, \etab_1, \etab_2)\pr_0(\thetab_1' \mid \thetab_2, \etab_1, \etab_2) \\
  &~~~\times \pr_0(\thetab_2' \mid \thetab_1', \etab_1, \etab_2)\pr_0(\etab_1' \mid \thetab_1', \thetab_2', \etab_2) \\
  &~~~\times \pr_0(\etab_2' \mid \thetab_1', \thetab_2', \etab_1')\intd \thetab_2 \intd \etab_1 \intd \etab_2 \\
 & = \iint \pr_0(\thetab_1' , \etab_1, \etab_2)\pr_0(\thetab_2' \mid \thetab_1', \etab_1, \etab_2)  \\
 & ~~~ \times \pr_0(\etab_1' \mid \thetab_1', \thetab_2', \etab_2)\pr_0(\etab_2' \mid \thetab_1', \thetab_2', \etab_1')\intd \etab_1 \intd \etab_2 \\
  & = \int \pr_0(\thetab_2',  \thetab_1', \etab_2) \pr_0(\etab_1' \mid \thetab_1', \thetab_2', \etab_2) \\
  & ~~~ \times \pr_0(\etab_2' \mid \thetab_1', \thetab_2', \etab_1')\intd \etab_2 \\
 & = \pr_0(\etab_1', \thetab_1', \thetab_2')\pr_0(\etab_2' \mid \thetab_1', \thetab_2', \etab_1'), \\
 & = \pr_0(\thetab_1', \thetab_2', \etab_1', \etab_2'),
\end{align*}
as required. Now, this vanilla sampler does not mix well due to the correlation a posteriori between $\etab_i$ and $\thetab_i$, $i = 1,2$ \citep{Knorr_2002}. Since our model is in a linear, Gaussian, setting, one might be tempted to instead use the following transition kernel:
\begin{align}
  q(\Thetab' \mid \Thetab) = & \pr_0(\thetab_1' \mid \thetab_2)\pr_0(\thetab_2' \mid \thetab_1') \times \nonumber \\
  & \pr_0(\etab_1' \mid \thetab_1', \thetab_2', \etab_2)\pr_0(\etab_2' \mid \thetab_1', \thetab_2', \etab_1').  \label{eq:q2}
 \end{align}
A similar treatment to the vanilla Gibbs case reveals that this only targets the correct distribution if $\pr_0(\thetab_1, \thetab_2, \etab_1, \etab_2) = \pr_0(\thetab_1, \thetab_2)\pr_0(\etab_1,\etab_2)$, which is almost certainly not the case in our spatial models. Therefore, \eqref{eq:q2} is not a target-preserving kernel. In our MCMC scheme this is important: updating $\thetab_k$ in blocks and subsequently updating $\etab_k$ in blocks will not yield samples from the posterior distribution.

A kernel which preserves the target can be constructed by updating $\etab_1$ and $\etab_2$ twice, with the intermediate quantities then discarded. As in Algorithm \ref{alg:spatial}, denote these intermediate quantities as $\etab_1^*$ and $\etab_2^*$, respectively, and consider the transition kernel
\begin{align*}
  q(\etab_1^*, \etab_2^*, \Thetab' \mid \Thetab) = & \pr_0(\etab_1^*, \thetab_1' \mid \thetab_2,  \etab_2)\pr_0(\etab_2^*, \thetab_2' \mid \thetab_1', \etab_1^*) \times \nonumber \\
  & \pr_0(\etab_1' \mid \thetab_1', \thetab_2', \etab_2^*)\pr_0(\etab_2' \mid \thetab_1', \thetab_2', \etab_1').  \label{eq:q3}
\end{align*}
This kernel preserves the target (posterior) distribution since
\begin{align*}
  \pr_1(\Thetab') & = \int\cdots\int \pr_0(\thetab_1, \thetab_2, \etab_1, \etab_2) \pr_0(\etab_1^*, \thetab_1' \mid \thetab_2,  \etab_2) \\
  & ~~~ \times \pr_0(\etab_2^*, \thetab_2' \mid \thetab_1', \etab_1^*)\pr_0(\etab_1' \mid \thetab_1', \thetab_2', \etab_2^*) \\
  & ~~~ \times \pr_0(\etab_2' \mid \thetab_1', \thetab_2', \etab_1')\intd \thetab_1\intd \thetab_2 \intd \etab_1 \intd \etab_2 \intd \etab_1^* \intd \etab_2^*  \\
  & = \iiiint \pr_0(\etab_1^*, \thetab_1' , \thetab_2,  \etab_2) \\
  & ~~~ \times \pr_0(\etab_2^*, \thetab_2' \mid \thetab_1', \etab_1^*)\pr_0(\etab_1' \mid \thetab_1', \thetab_2', \etab_2^*) \\
  & ~~~ \times \pr_0(\etab_2' \mid \thetab_1', \thetab_2', \etab_1')\intd \thetab_2 \intd \etab_2 \intd \etab_1^* \intd \etab_2^* \\
  & = \iint \pr_0(\etab_2^*, \thetab_2',  \thetab_1', \etab_1^*)\pr_0(\etab_1' \mid \thetab_1', \thetab_2', \etab_2^*) \\
  & ~~~ \times \pr_0(\etab_2' \mid \thetab_1', \thetab_2', \etab_1') \intd \etab_1^* \intd \etab_2^* \\
& = \int \pr_0(\etab_1', \thetab_1', \thetab_2', \etab_2^*)\pr_0(\etab_2' \mid \thetab_1', \thetab_2', \etab_1') \intd \etab_2^* \\
& =  \pr_0(\thetab_1', \thetab_2', \etab_1', \etab_2'),
\end{align*}
as required.

\section{\change{Simulation experiment illustrating the benefit of alternating tilings}}\label{app:sim_blocking}

\begin{algorithm-float}[t!]
\begin{framed}
\ifSTCO \color{blue} \fi
\begin{algorithm}
\label{alg:Sampler1} \textbf{Sampler 1 (Fixed tiling)}

\vspace{0.1in}

\noindent For $i = 1,\dots,N$
\begin{enumerate}
  \itemsep0em
  \item  Sample $\etab^{\curlyT_{11}} \mid \etab^{\curlyT_{12}}$
  \item  Sample $\etab^{\curlyT_{12}} \mid \etab^{\curlyT_{11}}$
\end{enumerate}
\end{algorithm}
\end{framed}

\begin{framed}
\begin{algorithm}
  \label{alg:Sampler2} \textbf{Sampler 2 (Alternating tilings)}
\vspace{0.1in}

\noindent For $i = 1,\dots,N$
\begin{enumerate}
  \itemsep1em
  \item If $i$ is odd:
  \begin{enumerate}
  \item  Sample $\etab^{\curlyT_{11}} \mid \etab^{\curlyT_{12}}$
  \item  Sample $\etab^{\curlyT_{12}} \mid \etab^{\curlyT_{11}}$
  \end{enumerate}
\item If $i$ is even:
  \begin{enumerate}
    \item  Sample $\etab^{\curlyT_{21}} \mid \etab^{\curlyT_{22}}$
    \item  Sample $\etab^{\curlyT_{22}} \mid \etab^{\curlyT_{21}}, \etab^{\curlyT_{23}}$
    \item  Sample $\etab^{\curlyT_{23}} \mid \etab^{\curlyT_{22}}$
      \end{enumerate}
\end{enumerate}
\end{algorithm}
\end{framed}
\end{algorithm-float}

\begin{figure*}[t!]
	\begin{center}
	  \includegraphics[width=3in]{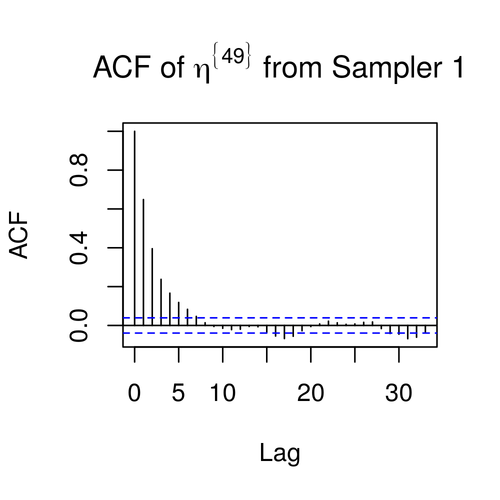}
          \includegraphics[width=3in]{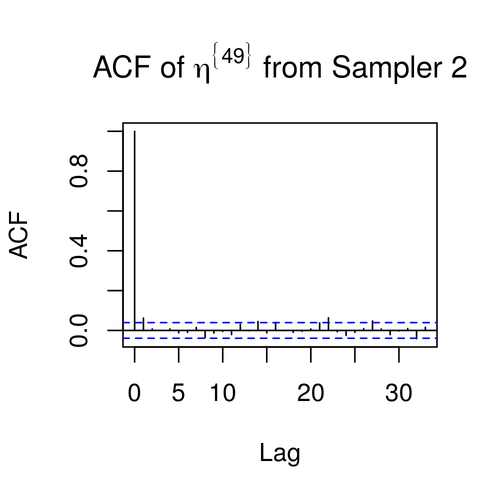}            
          \caption{\change{Empirical auto-correlation functions (ACF) from the Markov chain corresponding to $\etab^{\{49\}}$ from Sampler 1 (left) and Sampler 2 (right), respectively.} \label{fig:ACFs}}
	\end{center}
\end{figure*}

Consider a GMRF $\etab \sim \Gau(\zerob, \Qmat^{-1})$, where $\Qmat$ is the sparse tridiagonal matrix
\begin{equation}\label{eq:Q}
\Qmat = \frac{1}{\sigma^2_{v}}
\begin{pmatrix}
1     & -\phi     &           &      &    &  \\
-\phi & 1 + \phi^2 & -\phi      &      &    &  \\
     &   -\phi    & 1 + \phi^2 & -\phi &    &  \\
 &     &  \ddots    & \ddots& \ddots  &  \\
 &     &      & -\phi&  1 + \phi^2      & -\phi \\
      &          &           &      &   -\phi & 1
\end{pmatrix},
\end{equation}
where omitted entries are zero, $\phi$ is a length-scale parameter, and $\sigma^2_v$ a variance parameter. We consider the case where $\etab \in \mathbb{R}^n$ with $n = 99$, and consider the following tilings,
\begin{align*}
  \curlyT_{11} \equiv \{1,\dots, 49\},~\,  \qquad\qquad & \curlyT_{21} \equiv \{1,\dots, 33\}, \\
  \curlyT_{12} \equiv \{50,\dots, 99\},  \qquad\qquad & \curlyT_{22} \equiv \{34,\dots, 66\}, \\
                                                     & \curlyT_{23} \equiv \{67,\dots, 99\}.  
\end{align*}
Here, we compare Markov chain behaviour when sampling $N$ times from this distribution using the two samplers, Sampler 1 and Sampler 2, detailed in Algorithms~\ref{alg:Sampler1} and \ref{alg:Sampler2}, respectively. Sampler 1 is a blocked Gibbs sampler which samples $\etab$ using only $\curlyT_{11}$ and $ \curlyT_{12}$, while Sampler 2 changes the tiling used, $\{\curlyT_{11},\curlyT_{12}\}$ or $\{\curlyT_{21},\curlyT_{22},\curlyT_{23}\}$, at each iteration.

In our simulation experiment we simulated $\etab$ using $\phi = 0.9$ and $\sigma^2_v = 0.2$, and let $N = 10000$. We then generated two Markov chains, one using Sampler 1 and another using Sampler 2, and for each chain took the last 5000 samples and applied a thinning factor of 2. In Fig.~\ref{fig:ACFs} we show the empirical auto-correlation functions computed from the trace plots of $\etab^{\{49\}}$ from both samplers. We see that samples of $\etab^{\{49\}}$ from Sampler 1 are highly correlated due to the proximity of this variable to the tiling boundary. Samples of $\etab^{\{49\}}$ from Sampler 2 are virtually uncorrelated. Hence, a system of shifting tiles in a Gibbs sampler for spatial GMRFs (as done in Algorithm~\ref{alg:spatial}) can virtually eliminate any auto-correlation that may appear due to tile boundaries. Note that a thinning factor greater than the number of tilings needs to be used to effectively remove any auto-correlation.

\section{\change{Simulation experiment illustrating the sensitivity of the predictions to the chosen basis functions}}\label{app:sim_coarseness}

\begin{figure*}[!t]
	\begin{center}
		\includegraphics[width=0.9\textwidth]{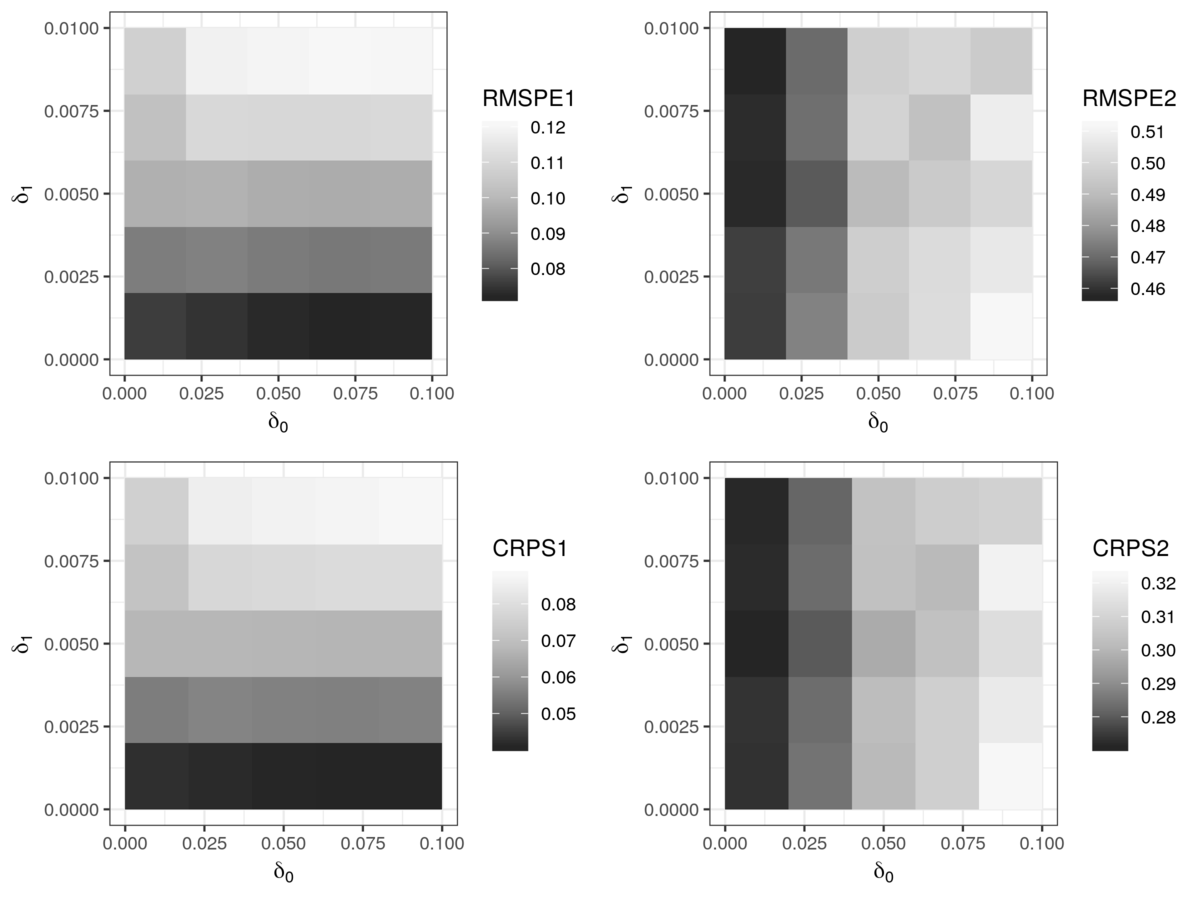}  
		\caption{\change{RMSPE (top panels) and CRPS (bottom panels) corresponding to $\Zvec_v^{(1)}$ (left panels) and $\Zvec_v^{(2)}$ (right panels) for varying basis-function widths ($\delta_0$ and $\delta_1$) in the 1D experiment.}} \label{fig:RMSPE_CRPS}
	\end{center}
\end{figure*}

In this section we conduct a simulation experiment that demonstrates the effect of a coarse discretisation, and the corresponding basis-function representation, on the prediction performance of the multi-scale model. The experiment is done in a one-dimensional, two-scale-process setting.

Consider a 1D Gaussian process on $D = [0,1]$, which has as covariance function a sum of two exponential covariance functions, $C_0(\cdot)$ and $C_1(\cdot)$. The exponential covariance functions have as parameters the variances $\sigma^2_k$ and ranges $\tau_k$, $k = 0, 1$, and are given by
$$
C_k(\hvec) = \sigma^2_k\exp(-\|\hvec\| / \tau_k), \quad k = 0,1.
$$

We model the process of interest $Y(\cdot)$ as a sum of the two processes $Y_k(\cdot) = \avec_k(\cdot)^\top\etab_k, k = 0, 1$, where $\avec_k(\cdot)$ are basis functions and $\etab_k$ are basis-function coefficients. Now, let the basis functions $\avec_k(\cdot), k = 0, 1,$ be piecewise constants on a regular partitioning of $D$. The basis functions have width $\delta_0$ and $\delta_1$, respectively, where $\delta_1 < \delta_0$. We are interested in what the effect of a poor choice for $\delta_0$ and $\delta_1$ is when predicting $Y(\cdot)$ from noisy observations $\Zvec$.

If $\delta_k \rightarrow 0$, then the $k$th scale of the original process is reconstructed exactly if we let $\etab_k \sim \Gau(\zerob, \Qmat_k^{-1}), k = 0,1,$ where $\Qmat_k$ is the (sparse) tridiagonal matrix given by \eqref{eq:Q} with $\sigma^2_v$ replaced with $\sigma^2_{v,k}$ and $\phi$ replaced with $\phi_k$. Here, $\phi_k =  \exp(-\delta_k / \tau_k)$ and $\sigma^2_{v,k} = \sigma^2_k(1 - \phi_k^2)$. In a simulation environment where we have access to $\sigma^2_k, \tau_k, k = 0,1$, we can therefore obtain accurate GMRF representations for our processes at the individual scales for when $\delta_k$ is small. We can then also see what happens as $\delta_k$ grows (this corresponds to coarsening a triangulation in 2D). 

In our simulation environment we fixed $\tau_0 = 0.4, \tau_1 = 0.04, \sigma^2_0 = 1$ and $\sigma^2_1 = 0.05$ and conducted 100 Monte Carlo simulations, where in each simulation we did the following:
\begin{enumerate}
	\item Randomly established an `unobserved' region in $D$, $D_{\gap}$ say, where $|D_{\gap}| = 0.2$.
	
	\item Generated 1100 observations on $D$, with 1000 in $D \backslash D_{\gap}$ and 100 in $D_{\gap}$ with measurement-error variance $\sigma^2_\epsilon = 0.0002$. Five hundred of those in $D\backslash D_{\gap}$ were used as training data $\Zvec$, the remaining 500 in $D \backslash D_{\gap}$ as validation data $\Zvec_v^{(1)}$, and those 100 in $D_{\gap}$ as validation data $\Zvec_v^{(2)}$.
	
	\item For various values of $\delta_0$ and $\delta_1$, we constructed $\Qmat_0$ and $\Qmat_1$ according to \eqref{eq:Q} and used these, the true measurement-error variance, and $\Zvec$, to predict $Y(\cdot)$ at all validation data locations.
	\item Computed the RMSPE and CRPS at the validation locations.
\end{enumerate}

Each Monte Carlo simulation provided us with an RMSPE and a CRPS corresponding to a combination of $\{\delta_0, \delta_1\}$. We then averaged over the 100 simulations to provide averaged RMSPEs and CRPSs corresponding to each combination of $\{\delta_0, \delta_1\}$. This experiment allows us to analyse the detrimental effect of a large $\delta_0$ or $\delta_1$ on our predictions.

The results from this experiment are summarised in Fig.~\ref{fig:RMSPE_CRPS}.  The figure clearly shows that the RMSPE considerably increases in regions where data is dense ($\Zvec_v^{(1)}$, left panels) and $\delta_1$ is large; on the other hand $\delta_0$ does not have much of an effect in these regions. The situation is reversed in regions where data is missing in large contiguous blocks ($\Zvec_v^{(2)}$, right panels). Here $\delta_1$ does not play a big role while $\delta_0$ does. When doing simple kriging (with the exact model) the mean RMSPEs were 0.07 and 0.41, respectively, while the mean CRPSs were 0.033 and 0.24, respectively, which are relatively close to what was obtained with the smallest values we chose for  $\delta_0$ and $\delta_1$. Therefore, the way in which we discretise both scales are important, and both $\delta_0$ and $\delta_1$ should be made as small as needed for their respective scale; in this experiment $\delta_1 = 0.001$ and $\delta_0 = 0.01$ are suitable choices. In practice, the coarseness of the grid (or triangulation in 2D) will be determined through computational considerations. Fortunately, we see that predictive performance deteriorates at an reasonably slow rate as the discretisations get coarser and coarser. A detailed analysis taking into account convergence rates of finite-element approximations might be needed for an in-depth analysis.

\end{document}